\documentclass[11pt]{article}

\usepackage[letterpaper,margin=1in]{geometry}
\usepackage[T1]{fontenc}
\usepackage[utf8]{inputenc}
\usepackage{newtxtext,newtxmath}
\usepackage{amsmath}
\usepackage{graphicx}
\usepackage{hyperref}
\usepackage{url}
\usepackage{xurl}
\usepackage{array}
\usepackage{longtable}
\hypersetup{hidelinks}

\providecommand{\tightlist}{%
  \setlength{\itemsep}{0pt}\setlength{\parskip}{0pt}}

\makeatletter
\renewcommand{\fnum@figure}{\textbf{Figure \thefigure}}
\renewcommand{\fnum@table}{\textbf{Table \thetable}}
\makeatother

\title{\textbf{Tidal Heating of the Lunar Magma Ocean: Reconciling an Old Moon with a Young Solidification}}
\author{Wenhao Zhao\textsuperscript{1}, Harriet Lau\textsuperscript{1}, Stephen Parman\textsuperscript{1}, James W. Head III\textsuperscript{1}}
\date{}

\begin{document}

\maketitle

\begin{center}
\textsuperscript{1}Department of Earth, Environmental and Planetary Sciences, Brown University, Providence, RI 02912, USA

Correspondence: \href{mailto:wenhao_zhao@brown.edu}{wenhao\_zhao@brown.edu}
\end{center}

\hypertarget{abstract}{%
\section*{Abstract}\label{abstract}}

The timing of the Moon's formation is fundamental to understanding the early Earth--Moon system. Ages of lunar magma ocean (LMO) crystallization have long been regarded as a key proxy for that event. Yet returned lunar sample ages cluster near the relatively young age of \textasciitilde4.35 billion years ago (Ga). These ages are commonly interpreted as recording either a young-Moon formation age or later thermal resetting. Here we show that, for an old Moon (\textgreater4.5 Ga), the \textasciitilde4.35 Ga age cluster can instead arise naturally from early LMO thermal evolution under Earth's tidal forcing. We identify tidal heating within a partially molten LMO as a major internal heat source. It offsets much of the early heat loss and maintains a long-lived high-energy state for \textgreater150 million years. As crystallization proceeded, this stable state was ultimately lost through the rapid collapse of tidal heating. The last stages of LMO solidification were compressed into a short interval near \textasciitilde4.35 Ga. The tidal heat source decouples Moon formation from final LMO solidification. As an outcome of LMO evolution, we predict asymmetric late-stage crystallization between the lunar nearside and farside, potentially linking tidally modulated LMO evolution to the long-term lunar dichotomy.

\hypertarget{introduction}{%
\section{Introduction}\label{introduction}}

The Moon is widely thought to have formed through a giant impact between the proto-Earth and a Mars-sized body\textsuperscript{1,2}. Constraining the timing of this event is essential for understanding the early thermal, dynamical, and chemical evolution of the Earth--Moon system. This requires a process that is genetically linked to Moon formation and early differentiation, and amenable to radiometric dating. Among candidate processes, crystallization of the lunar magma ocean (LMO), a global differentiation event initiated during or soon after the Moon-forming impact, was considered particularly well suited\textsuperscript{3--5}. During this process, dense mafic cumulates sank, buoyant plagioclase-rich cumulates rose and formed a ferroan anorthosite (FAN) crust, late-stage crystallization produced KREEP [potassium (K), rare earth elements (REE), and phosphorus (P)]-rich residual melts, and subsequent cumulate overturn and associated magmatism generated the so-called Mg-suite intrusions\textsuperscript{1,4,6,7}. These distinct reservoirs provide largely independent isotopic constraints on early lunar evolution.

Returned Apollo samples indicate that multiple lithologies associated with LMO differentiation and its subsequent evolution yield internally consistent ages across diverse isotopic systems. Collectively, these results define a prominent age cluster at $4.35 \pm 0.05$~Ga (Fig. 1A). This clustering is reflected in five principal observations: (1) a \textsuperscript{146}Sm-\textsuperscript{142}Nd isochron age for the formation of mare basalt source regions; (2) crystallization ages of the lunar primary crust recorded by FANs; (3) \textsuperscript{147}Sm-\textsuperscript{143}Nd and \textsuperscript{176}Lu-\textsuperscript{176}Hf model ages for KREEP-rich residual melts; (4) crystallization ages of Mg-suite lithologies interpreted as later intrusive products associated with cumulate overturn; and (5) a prominent \textasciitilde4.35 Ga peak in detrital zircons derived from the lunar crust\textsuperscript{3,5,8--10}.

The \textasciitilde4.35 Ga age cluster has two notable features. It is much younger than Solar System formation at \textasciitilde4.567 Ga\textsuperscript{11}, implying a delay of \textgreater150 Myr, and it is tightly concentrated, suggesting rapid termination rather than prolonged evolution. Two main interpretations have been proposed. In the ``young-Moon'' model, the Moon formed at \textasciitilde4.35 Ga and the cluster records rapid LMO crystallization shortly thereafter (Fig. 1B)\textsuperscript{5,8}. In the ``reset-Moon'' model, lunar formation and primary LMO solidification occurred earlier, and the cluster instead reflects later isotopic resetting by a thermal or magmatic event such as large-scale impact reheating or orbital-tidal reorganization (Fig. 1C)\textsuperscript{12}. Both interpretations, however, face major difficulties. Regarding the former interpretation, a \textasciitilde4.35 Ga Moon is inconsistent with recent Rb-Sr constraints from Apollo 16 ferroan anorthosite 60025, which require the Moon to be older than \textasciitilde4.49 Ga\textsuperscript{13}; initial Sr-isotope systematics of additional lunar ferroan anorthosites likewise imply Moon formation at \textasciitilde4.50 Ga\textsuperscript{14}; Hf-W systematics linked to lunar core formation similarly favor early Moon formation\textsuperscript{15}; and zircons approaching \textasciitilde4.5 Ga indicate that at least part of lunar differentiation predates \textasciitilde4.35 Ga\textsuperscript{10}. For the latter, the reset-Moon interpretation requires a later event near \textasciitilde4.35 Ga capable of resetting multiple isotopic systems without erasing early lunar records; both its timing and physical plausibility remain uncertain in current orbital and impact scenarios\textsuperscript{12,16--18}.

Despite their differences, both interpretations assume that a substantial degree of LMO solidification occurred relatively soon after Moon formation\textsuperscript{8,12}. To address the tensions outlined above, some studies have instead explored frameworks in which Moon formation and LMO crystallization are explicitly decoupled. Under this viewpoint, the duration of LMO solidification is extended by invoking mechanisms that reduce heat-loss efficiency, such as thermal insulation by an early crust or transient atmosphere\textsuperscript{4,19}. In these frameworks \textasciitilde4.35 Ga can represent the age of LMO shutdown, while the Moon itself formed earlier (Fig. 1D). However, existing work along these lines still struggles to explain the observed age cluster, because once rapid post-formation cooling is no longer assumed, products from different stages of LMO differentiation should record a spectrum of ages, rather than the tight age clustering that is observed. Therefore, the geological meaning of the \textasciitilde4.35 Ga age cluster and its implications for lunar formation and early evolution remain unresolved.

We propose that tidal heating of the partially solidified LMO may explain the geochronologic observations outlined above. This model is distinct from the interpretations discussed above: the Moon formed early (\textgreater4.5 Ga), remained partially molten for \textasciitilde150 Myr, and then abruptly completed the remaining LMO solidification at \textasciitilde4.35 Ga (Fig. 1A). By introducing tidal heating within the magma ocean as a major heat source during this interval, the net cooling of the LMO reflects the competition between heat loss and tidal heating rather than being overwhelmingly controlled by heat loss. This chronologically decouples lunar formation from LMO solidification and naturally produces a concentrated age cluster near \textasciitilde4.35 Ga. Results from our parameterized coupled orbital-thermal model of the early Moon show a long-lived high-dissipation buffered state followed by rapid cooling driven by the collapse of tidal heating through a rheological mechanism described below. It further suggests that tidally dominated LMO evolution may have produced the nearside-farside dichotomy and established hemispheric differences in crustal structure, geochemical and volatile enrichment, and deep-interior physical state that persisted as starting conditions for the Moon's subsequent billions of years of evolution, in directions broadly consistent with current observations\textsuperscript{20--29}.

\hypertarget{tidal-heating-decouples-moon-formation-from-lmo-solidification}{%
\section{Tidal Heating Decouples Moon Formation from LMO Solidification}\label{tidal-heating-decouples-moon-formation-from-lmo-solidification}}

Dating Moon formation through LMO solidification assumes that the thermal interval between these events was short\textsuperscript{5}. In most thermal-evolution models, that assumption follows naturally from the energy budget: vigorous convection dominates heat loss, whereas radiogenic heating is small and evolves slowly\textsuperscript{30,31}. During the magma-ocean stage, whole-Moon heat loss exceeds $10^3$~terawatts (TW)\textsuperscript{31}, whereas our calculations give only \textasciitilde1.8 TW of total radiogenic power at 4.5 Ga and \textasciitilde1.4 TW at 4.0 Ga (Supplementary Fig.~S2). This mismatch has long supported the view that LMO cooling was controlled mainly by heat loss.

We argue that this energy budget is incomplete because it neglects tidal heating within the magma ocean itself. Tidal heating converts orbital and rotational energy from the Earth--Moon system into lunar internal heat through periodic forcing and viscoelastic dissipation\textsuperscript{32--34}. Most previous studies of lunar tidal dissipation have focused either on a largely solid lunar interior\textsuperscript{33} or on specific later orbital-tidal events such as the \textasciitilde4.35~Ga Laplace plane transition\textsuperscript{12}, rather than on the tidal response of the LMO itself. Chen and Nimmo\textsuperscript{35} provided one example involving LMO dissipation, but treated it mainly in the context of early orbital evolution rather than its thermal and energetic effects on LMO solidification. Related problems have, however, been studied for other bodies in partially molten or magma-ocean-like regimes, including Io and tidally active exoplanets\textsuperscript{36--38}. During the LMO epoch, both ingredients favored strong dissipation: the Moon orbited much closer to Earth, so the tide-raising forcing was far stronger than today and the associated tidal heating scales steeply with Earth--Moon semi-major axis $a$, approximately as $a^{-6}$ under otherwise comparable conditions\textsuperscript{34,39}, and a partially molten interior could dissipate energy much more efficiently than either a fully liquid or fully solid Moon under comparable forcing\textsuperscript{36--38}. Because viscoelastic dissipation peaks when forcing and stress-relaxation timescales are comparable, crystallization-driven changes in melt fraction and viscosity define a high-dissipation window during LMO evolution (Supplementary Fig.~S3A,B), typically at intermediate melt fractions rather than either endmember limit\textsuperscript{40--42}. Whether crystal settling and convective re-entrainment, together with continued tidal heating, can maintain the partially molten crystal--melt mixture required for strong dissipation is discussed in Supplementary Section 3.7.

Our model therefore predicts three stages of LMO thermal evolution (Fig. 2). Stage 1 corresponds to the immediate post-formation interval (\textasciitilde4.50 Ga), when the magma ocean is too fluid to dissipate efficiently, so tidal heating remains weak and cooling proceeds rapidly. The detailed behavior of this earliest, highly molten stage is likely complex and depends on poorly constrained rheology and thermal structure. However, because the system evolves rapidly into the high-dissipation viscoelastic window (within \textless1 Myr, and specifically \textasciitilde0.14 Myr in the baseline case), these uncertainties have limited influence on the later evolution emphasized here. Here and below, the baseline case denotes a simplified orbital treatment with fixed eccentricity $e=0.05$ and a prescribed Earth--Moon semi-major-axis history $a(t)$ generated with constant $Q_{\mathrm{E}}/k_{2\mathrm{E}}=400$ (Supplementary Section 2.4); its location in the explored parameter space is marked in Supplementary Fig.~S6. The effects of varying these orbital assumptions are examined in Supplementary Section 3.2, and possible feedbacks of tidal dissipation on $e(t)$ and $a(t)$ are discussed in Supplementary Sections 3.5 and 3.6.

Stage 2 begins once temperature and melt fraction enter the viscoelastic window. Tidal dissipation then rises sharply to hundreds of terawatts in the baseline model, approaching the convective heat-loss rate. The system enters a long-lived buffered state in which tidal heating offsets a substantial fraction of surface heat loss. In a manner analogous to stable equilibria discussed in tidal-heating studies of Io and exoplanets\textsuperscript{38}, small perturbations in temperature or melt fraction are opposed by negative feedbacks that drive the system back toward the same high-dissipation, tidally buffered state. As the Moon recedes, this stable equilibrium migrates toward lower temperature and melt fraction, approaching the dissipation maximum near melt fractions of \textasciitilde30--50\%. By \textasciitilde4.35 Ga, baseline tidal heating still exceeds 30 TW, more than an order of magnitude larger than radiogenic heating (Fig. 2 and Supplementary Fig.~S6).

Across broad parameter variations, the same dynamical structure persists: an initial weak-dissipation cooling stage followed by a long tidally buffered state whose onset and duration vary among models\textsuperscript{38}. Tidal dissipation can therefore reshape the overall cooling history, decouple lunar formation from LMO solidification, and make an old Moon mechanically compatible with a younger \textasciitilde4.35 Ga shutdown age (Fig. 2 and Supplementary Fig.~S6).

\hypertarget{rapid-collapse-of-tidal-heating-produces-the-4.35-ga-age-cluster}{%
\section{Rapid Collapse of Tidal Heating Produces the \textasciitilde4.35 Ga Age Cluster}\label{rapid-collapse-of-tidal-heating-produces-the-4.35-ga-age-cluster}}

The strong concentration of ages near \textasciitilde4.35 Ga is readily explained by existing interpretations. In the young-Moon view, such concentration follows naturally because Moon formation occurs at \textasciitilde4.35 Ga and rapid cooling immediately generates clustered crystallization ages\textsuperscript{5,8}. In the reset-Moon model, the same concentration is instead attributed to a later resetting event and need not arise from primary LMO thermal evolution at all\textsuperscript{12}. In our framework, however, once Stage 2 succeeds in prolonging the partially molten state toward \textasciitilde4.35 Ga, the central question becomes how that delayed evolution ends rapidly enough to produce the observed clustering.

The buffered Stage-2 state cannot persist indefinitely (Fig. 2). As crystallization proceeds and the Moon continues to recede, the stable equilibrium migrates toward lower temperature and melt fraction until the system reaches the unstable equilibrium that bounds the high-dissipation branch, marking Stage 3 of the evolution. Unlike the stable equilibrium in Stage 2, small perturbations near this unstable equilibrium drive the system away from the tidally buffered state. Beyond this point, the feedback changes sign: a small decrease in temperature or melt fraction weakens tidal dissipation rather than restoring it. Net heat loss therefore rises, cooling accelerates, and that cooling further suppresses dissipation. The result is a positive feedback that rapidly collapses tidal heating. In this framework, the rapid terminal solidification is caused not by a sudden increase in heat loss, but by the collapse of the tidal-heating source that had previously buffered that heat loss. The corresponding evolution of the major heating and cooling terms in the baseline model is shown in Fig. 3. The remaining LMO solidification is therefore compressed into a relatively short interval rather than spread gradually through time.

Across the explored parameter space, model choices mainly shift when the system leaves the buffered stage, not the existence of the collapse itself. Once that stable equilibrium is lost, the remaining solidification is naturally compressed into a short interval, providing a simple explanation for the tightly clustered \textasciitilde4.35 Ga ages recorded by multiple LMO-related chronometers (Fig. 2 and Supplementary Fig.~S6).

\hypertarget{the-long-term-legacy-of-tidally-asymmetric-lmo-evolution}{%
\section{The Long-Term Legacy of Tidally Asymmetric LMO Evolution}\label{the-long-term-legacy-of-tidally-asymmetric-lmo-evolution}}

Terminal LMO crystallization established the thermal, chemical, and mechanical initial conditions from which the Moon evolved over the following billions of years. If tidal heating modulated late LMO evolution, that process should have left observable signatures in the present-day lunar crust, mantle, and hemispheric structure. This expectation is especially strong if nearside and farside LMO heating differed because they lay at different distances from Earth.

Such signatures may already exist. The nearside-farside dichotomy extends from crust to deep interior\textsuperscript{22}. GRAIL gravity and topography show that the farside crust is on average \textasciitilde15--20 km thicker than the nearside, and mare volcanism is overwhelmingly concentrated on the nearside\textsuperscript{20,21,43,44}. Geochemical and remote-sensing data indicate nearside enrichment in incompatible elements and Fe-Ti-rich materials, whereas farside highlands include more magnesian and primitive anorthositic compositions\textsuperscript{23,24,45}. Recent Chang'e-6 results further suggest a cooler, more depleted, and more volatile-poor farside mantle source\textsuperscript{26--28}, while tidal-response inversions point to a mechanically softer nearside mantle at depth\textsuperscript{29}. Together, these observations suggest that the final products of early differentiation were not hemispherically uniform.

Many mechanisms have been proposed to explain these asymmetries, including asymmetric accretion, mantle overturn, thermal evolution, radiogenic concentration, and earlier asymmetric-crystallization frameworks.\textsuperscript{4,25,46--50} In our framework, however, a small nearside-farside asymmetry is built into the LMO energy budget through higher-order finite-distance corrections to the Earth-raised tide, because the nearside and farside lie one lunar radius closer to, or farther from, Earth. To estimate the first-order consequences, we treat the two hemispheres as small radial offsets of $\pm\Delta R$ relative to the Earth--Moon separation ($\Delta R = R_{\mathrm{M}}$) and apply the resulting nearside-farside forcing contrast as a hemispheric weighting on tidal dissipation (Supplementary Section 2.5). Across most of Stage 2, the modeled temperature contrast is small ($<5\,^\circ$C), but it grows sharply near the Stage 3 collapse because the farside reaches the unstable equilibrium earlier than the nearside (Fig. 4). In our explored parameter space, peak contrasts reach \textasciitilde20--100~$^\circ$C for \textasciitilde10 to \textgreater100 Myr; for cliff ages of 4.4--4.3 Ga, the contrast is \textasciitilde20--60~$^\circ$C over \textasciitilde10--20 Myr (Supplementary Fig.~S6B,C). Although this treatment is zero-dimensional and therefore only semi-quantitative, it indicates that the final stage of LMO evolution need not have been hemispherically synchronous.

That late asymmetry provides a simple link between tidally buffered LMO evolution and the Moon we observe today (Fig. 5). If the farside entered terminal crystallization earlier, it should on average preserve older and less evolved anorthositic crust, whereas the warmer nearside should retain more evolved residual components. This framework therefore predicts hemispheric contrasts in crustal thickness, age, and composition; preferential nearside enrichment in KREEP, Fe-Ti oxides, water, and other incompatible components; and long-wavelength differences in deep-interior temperature, partial melting, and shear modulus. Several of these trends are already broadly consistent with existing observations of crustal structure, chemical enrichment, farside mantle depletion, and deep-interior asymmetry\textsuperscript{20,21,23--29}, but direct farside age constraints and fully coupled 2D-3D models are still needed to test whether asymmetric tidal heating can quantitatively account for the lunar dichotomy.

\hypertarget{methods}{%
\section{Methods}\label{methods}}

We model the lunar magma ocean (LMO) as a homogeneous, well-mixed silicate layer whose mean thermal evolution is governed by the competition among tidal dissipation, radiogenic heating, core--mantle boundary heat exchange and convective heat loss. The time evolution of the mean LMO temperature is obtained from an energy-balance equation in which tidal heating and radiogenic decay add energy to the reservoir, convective transport removes energy through the surface, and basal exchange couples the LMO to a well-mixed lunar core. Latent-heat buffering during crystallization is incorporated through an effective heat capacity. We track both the LMO temperature and the core temperature through time. The model is intentionally zero-dimensional and parameterized so that the dominant controls on the thermal budget can be isolated across the poorly constrained earliest stages of lunar evolution.

Radiogenic heating is computed from the decay of the four dominant long-lived heat-producing isotopes (\textsuperscript{232}Th, \textsuperscript{238}U, \textsuperscript{235}U and \textsuperscript{40}K) using present-day concentrations and decay constants. Convective heat loss is treated with parameterized upper and lower thermal boundary layers, following Nusselt--Rayleigh scaling. The upper boundary layer controls heat loss from the LMO to the surface, and the lower boundary layer controls heat exchange across the core--mantle boundary. Surface temperature is solved from a gray-body balance between absorbed solar flux and interior heat release. This treatment captures the dominant thermal terms while avoiding assumptions about spatially resolved mantle circulation that are not presently constrained.

Tidal dissipation within the LMO is calculated with a viscoelastic Love-number formalism in which the global tidal power depends on the negative imaginary part of the degree-2 Love number, the Earth--Moon distance, mean motion and eccentricity. We adopt a Maxwell rheology for the partially molten LMO and parameterize melt fraction, shear modulus and shear viscosity as functions of temperature. The viscosity law is segmented between the solidus, a critical rheological transition and the liquidus so that dissipation peaks in the transitional window where forcing and stress-relaxation timescales are comparable. This is the key regime that permits strong tidal heating in a partially molten Moon. In the baseline calculation, the effective dissipating volume corresponds to the silicate shell above an undissipative core, scaled by the melt fraction as a proxy for the connected crystal--melt region able to participate in tidal dissipation.

Orbital forcing is prescribed through a simplified Earth-dominated recession history. In the baseline case, we assume synchronous lunar rotation, a constant eccentricity of $e=0.05$, an initial Earth--Moon distance of $3.0\,R_{\mathrm{E}}$, and a constant terrestrial dissipation ratio $Q_{\mathrm{E}}/k_{2\mathrm{E}}=400$, equivalent to $Q_{\mathrm{E}}=120$ and $k_{2\mathrm{E}}=0.3$. This produces an Earth--Moon distance history consistent with previous work and places the Moon at $\sim 29.4\,R_{\mathrm{E}}$ near 4.35~Ga in the baseline run. We initialize the LMO and core at $T_{m0}=3000$~K and $T_{c0}=3100$~K; supplementary sensitivity tests show that these starting values affect only the brief earliest transient and do not materially change the later buffered stage or cliff age.

To estimate nearside--farside asymmetry, we extend the symmetric model with a finite-distance correction to the Earth-raised tide. The nearside and farside are treated as radial offsets of $\pm R_{\mathrm{M}}$ from the lunar center, which introduces a small hemispheric difference in the tide-raising acceleration when the early Moon is close to Earth. We convert this difference into multiplicative corrections to the local tidal power and evolve two otherwise identical zero-dimensional thermal columns for the nearside and farside. This provides a first-order estimate of hemispheric differences in mantle temperature and cliff age without invoking pre-existing lateral heterogeneity.

Further details of the model formulation, governing equations, parameter values and expanded sensitivity analyses are provided in the Supplementary Materials.

\hypertarget{references}{%
\section*{References}\label{references}}

\begin{enumerate}
\def\labelenumi{\arabic{enumi}.}
\tightlist
\item
  Warren, P. H. The magma ocean concept and lunar evolution. Annual Review of Earth and Planetary Sciences 13, 201--240 (1985).
\item
  Canup, R. M. Simulations of a late lunar-forming impact. Icarus 168, 433--456 (2004).
\item
  Carlson, R. W. \& Lugmair, G. W. The age of ferroan anorthosite 60025: oldest crust on a young Moon? Earth and Planetary Science Letters 90, 119--130 (1988).
\item
  Elkins-Tanton, L. T., Burgess, S. \& Yin, Q.-Z. The lunar magma ocean: Reconciling the solidification process with lunar petrology and geochronology. Earth and Planetary Science Letters 304, 326--336 (2011).
\item
  Borg, L. E. \& Carlson, R. W. The evolving chronology of Moon formation. Annual Review of Earth and Planetary Sciences 51, 25--52 (2023).
\item
  Wood, J. A., Dickey, J. S. Jr., Marvin, U. B. \& Powell, B. N. Lunar anorthosites. Science 167, 602--604 (1970).
\item
  Schmidt, M. W. \& Kraettli, G. Experimental crystallization of the lunar magma ocean, initial selenotherm and density stratification, and implications for crust formation, overturn and the bulk silicate Moon composition. Journal of Geophysical Research: Planets 127, e2022JE007187 (2022).
\item
  Borg, L. E., Connelly, J. N., Boyet, M. \& Carlson, R. W. Chronological evidence that the Moon is either young or did not have a global magma ocean. Nature 477, 70--72 (2011).
\item
  Borg, L. E., Gaffney, A. M. \& Shearer, C. K. A review of lunar chronology revealing a preponderance of 4.34--4.37 Ga ages. Meteoritics \& Planetary Science 50, 715--732 (2015).
\item
  Barboni, M. et al.~Early formation of the Moon 4.51 billion years ago. Science Advances 3, e1602365 (2017).
\item
  Connelly, J. N. et al.~The absolute chronology and thermal processing of solids in the solar protoplanetary disk. Science 338, 651--655 (2012).
\item
  Nimmo, F., Kleine, T. \& Morbidelli, A. Tidally driven remelting around 4.35 billion years ago indicates the Moon is old. Nature 636, 598--602 (2024).
\item
  Yobregat, E., Fitoussi, C. \& Bourdon, B. Rb-Sr constraints on the age of Moon formation. Icarus 420, 116164 (2024).
\item
  Schneider, J. M. \& Kleine, T. The age and early evolution of the Moon revealed by the Rb-Sr systematics of lunar ferroan anorthosites. Earth and Planetary Science Letters 669, 119592 (2025).
\item
  Thiemens, M. M. et al.~Early Moon formation inferred from hafnium-tungsten systematics. Nature Geoscience 12, 696--700 (2019).
\item
  Cuk, M., Lock, S. J., Stewart, S. T. \& Hamilton, D. P. Tidal evolution of the Earth--Moon system with a high initial obliquity. The Planetary Science Journal 2, 147 (2021).
\item
  Farhat, M., Auclair-Desrotour, P., Bou\'{e}, G. \& Laskar, J. The resonant tidal evolution of the Earth--Moon distance. Astronomy \& Astrophysics 665, L1 (2022).
\item
  Su, B. et al.~South Pole-Aitken massive impact 4.25 billion years ago revealed by Chang'e-6 samples. National Science Review 12, nwaf103 (2025).
\item
  Lebrun, T. et al.~Thermal evolution of an early magma ocean in interaction with the atmosphere. Journal of Geophysical Research: Planets 118, 1155--1176 (2013).
\item
  Garrick-Bethell, I., Nimmo, F. \& Wieczorek, M. A. Structure and formation of the lunar farside highlands. Science 330, 949--951 (2010).
\item
  Wieczorek, M. A. et al.~The crust of the Moon as seen by GRAIL. Science 339, 671--675 (2013).
\item
  Andrews-Hanna, J. C. et al.~The structure and evolution of the lunar interior. Reviews in Mineralogy and Geochemistry 89, 243--292 (2023).
\item
  Jolliff, B. L. et al.~Major lunar crustal terranes: Surface expressions and crust-mantle origins. Journal of Geophysical Research: Planets 105, 4197--4216 (2000).
\item
  Ohtake, M. et al.~Asymmetric crustal growth on the Moon indicated by primitive farside highland materials. Nature Geoscience 5, 384--388 (2012).
\item
  Wieczorek, M. A. \& Phillips, R. J. The ``Procellarum KREEP Terrane'': Implications for mare volcanism and lunar evolution. Journal of Geophysical Research: Planets 105, 20417--20430 (2000).
\item
  He, H. et al.~Water abundance in the lunar farside mantle. Nature 643, 366--370 (2025).
\item
  He, S. et al.~A relatively cool lunar farside mantle inferred from Chang'e-6 basalts and remote sensing. Nature Geoscience 18, 1103--1108 (2025).
\item
  Zhou, Q. et al.~Ultra-depleted mantle source of basalts from the South Pole-Aitken basin. Nature 643, 371--375 (2025).
\item
  Park, R. S. et al.~Thermal asymmetry in the Moon's mantle inferred from monthly tidal response. Nature 641, 1188--1192 (2025).
\item
  Sahijpal, S. \& Goyal, V. Thermal evolution of the early Moon. Meteoritics \& Planetary Science 53, 2193--2211 (2018).
\item
  Solomatov, V. Magma oceans and primordial mantle differentiation. In Treatise on Geophysics, 2nd edn, Vol. 9, 81--104 (Elsevier, 2015).
\item
  Kopal, Z. Gravitational heating of the Moon. Icarus 1, 412--421 (1962).
\item
  Peale, S. J. \& Cassen, P. Contribution of tidal dissipation to lunar thermal history. Icarus 36, 245--269 (1978).
\item
  Murray, C. D. \& Dermott, S. F. Solar System Dynamics (Cambridge University Press, 1999).
\item
  Chen, E. M. A. \& Nimmo, F. Tidal dissipation in the lunar magma ocean and its effect on the early evolution of the Earth--Moon system. Icarus 275, 132--142 (2016).
\item
  Segatz, M., Spohn, T., Ross, M. N. \& Schubert, G. Tidal dissipation, surface heat flow, and figure of viscoelastic models of Io. Icarus 75, 187--206 (1988).
\item
  Henning, W. G., O'Connell, R. J. \& Sasselov, D. D. Tidally heated terrestrial exoplanets: viscoelastic response models. The Astrophysical Journal 707, 1000--1015 (2009).
\item
  Renaud, J. P. \& Henning, W. G. Increased tidal dissipation using advanced rheological models: Implications for Io and tidally active exoplanets. The Astrophysical Journal 857, 98 (2018).
\item
  Kaula, W. M. Tidal dissipation by solid friction and the resulting orbital evolution. Reviews of Geophysics 2, 661--685 (1964).
\item
  Arzi, A. A. Critical phenomena in the rheology of partially melted rocks. Tectonophysics 44, 173--184 (1978).
\item
  Harada, Y. et al.~Strong tidal heating in an ultralow-viscosity zone at the core-mantle boundary of the Moon. Nature Geoscience 7, 569--572 (2014).
\item
  Bierson, C. J. The impact of rheology model choices on tidal heating studies. Icarus 414, 116026 (2024).
\item
  Head, J. W. III \& Wilson, L. Lunar mare volcanism: Stratigraphy, eruption conditions, and the evolution of secondary crusts. Geochimica et Cosmochimica Acta 56, 2155--2175 (1992).
\item
  Zuber, M. T. et al.~Gravity field of the Moon from the Gravity Recovery and Interior Laboratory (GRAIL) mission. Science 339, 668--671 (2013).
\item
  Takeda, H. et al.~Magnesian anorthosites and a deep crustal rock from the farside crust of the Moon. Earth and Planetary Science Letters 247, 171--184 (2006).
\item
  Wasson, J. T. \& Warren, P. H. Contribution of the mantle to the lunar asymmetry. Icarus 44, 752--771 (1980).
\item
  Arai, T., Takeda, H., Yamaguchi, A. \& Ohtake, M. A new model of lunar crust: asymmetry in crustal composition and evolution. Earth, Planets and Space 60, 433--444 (2008).
\item
  Jutzi, M. \& Asphaug, E. Forming the lunar farside highlands by accretion of a companion moon. Nature 476, 69--72 (2011).
\item
  Laneuville, M., Wieczorek, M. A., Breuer, D. \& Tosi, N. Asymmetric thermal evolution of the Moon. Journal of Geophysical Research: Planets 118, 1435--1452 (2013).
\item
  Quillen, A. C., Martini, L. \& Nakajima, M. Near/far side asymmetry in the tidally heated Moon. Icarus 329, 182--196 (2019).
\item
  Taylor, S. R. Lunar and terrestrial crusts: a contrast in origin and evolution. Physics of the Earth and Planetary Interiors 29, 233--241 (1982).
\item
  Arevalo Jr., R., McDonough, W. F. \& Luong, M. The K/U ratio of the silicate Earth: Insights into mantle composition, structure and thermal evolution. Earth and Planetary Science Letters 278, 361--369 (2009).
\end{enumerate}

\section*{Acknowledgements}

We thank Greg Hirth, Reid Cooper, Christian Huber, and Marc Parmentier for helpful discussions.

This work was supported by the LunaSCOPE NASA SSERVI node (grant 80NSSC23M0161).

\clearpage
\begin{center}
\includegraphics[width=1.00\textwidth,height=0.49\textheight,keepaspectratio]{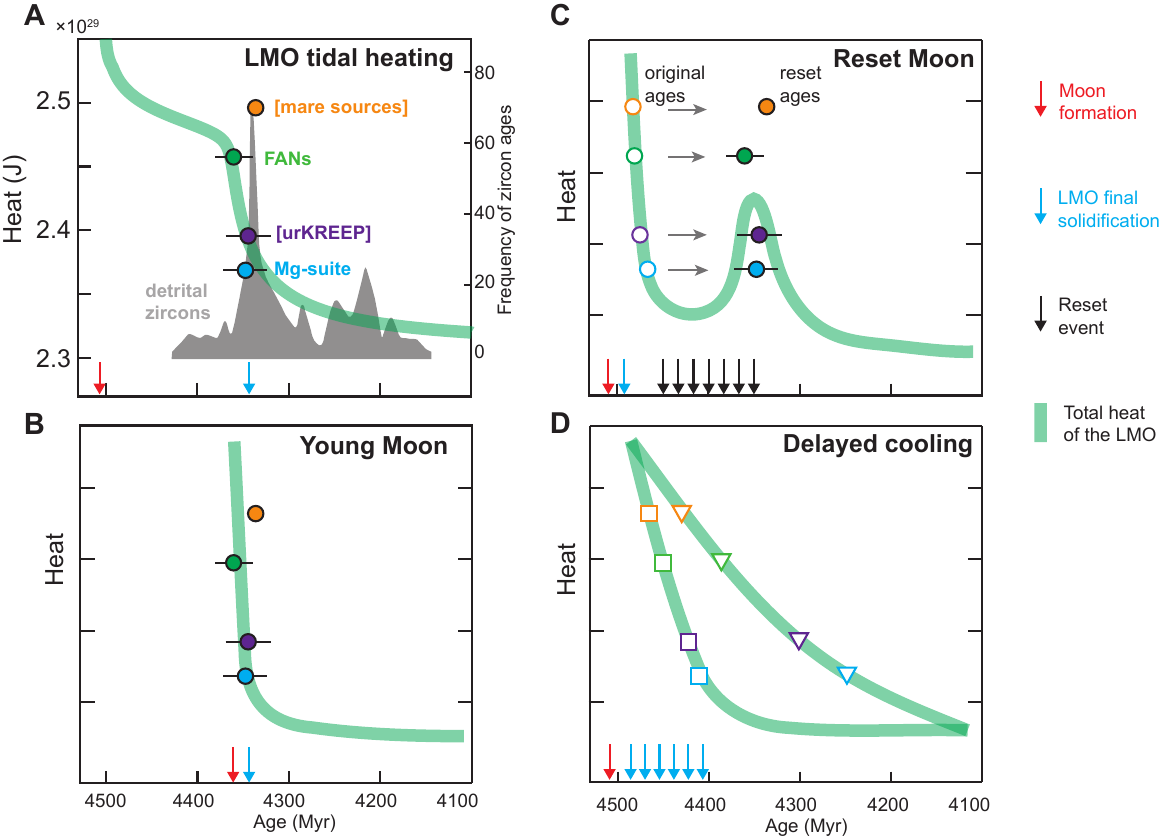}
\end{center}
\hypertarget{fig.-1.-competing-interpretations-of-the-4.35-ga-age-cluster-and-the-role-of-tidal-heating.}{%
\subsection*{Fig. 1. Four paradigms linking Moon formation, LMO thermal evolution, and the \textasciitilde4.35 Ga lunar age cluster.}\label{fig.-1.-competing-interpretations-of-the-4.35-ga-age-cluster-and-the-role-of-tidal-heating.}}

\noindent (A) Tidal-heating interpretation of this work. Colored circles mark representative ages from LMO-related lithologies and reservoirs, and the gray shaded distribution shows the relative frequency of detrital-zircon ages (modified from Borg and Carlson\textsuperscript{5}). The green curve denotes the modeled relative LMO heat content from the baseline net energy budget. Red arrows mark Moon formation, blue arrows mark terminal LMO solidification, and black arrows mark later resetting events. In this framework, tidal heating maintains a long-lived high-energy state and allows an old Moon to remain partially molten until rapid LMO shutdown near \textasciitilde4.35 Ga. (B) Young-Moon interpretation, in which Moon formation and rapid LMO cooling both occur near \textasciitilde4.35 Ga (after Borg et al.\textsuperscript{8}). (C) Reset-Moon interpretation, in which LMO-related units formed earlier but were subsequently reset by a later event; although invoked to explain the \textasciitilde4.35 Ga age concentration, that resetting need not have occurred exactly at \textasciitilde4.35 Ga (after Nimmo et al.\textsuperscript{12}). (D) Delayed-cooling interpretation driven by reduced heat loss alone; this can postpone LMO shutdown, but tends to spread ages across time rather than reproduce the observed cluster (after Elkins-Tanton et al.\textsuperscript{4} and Lebrun et al.\textsuperscript{19}).

\clearpage
\begin{center}
\includegraphics[width=0.82\textwidth,height=0.41\textheight,keepaspectratio]{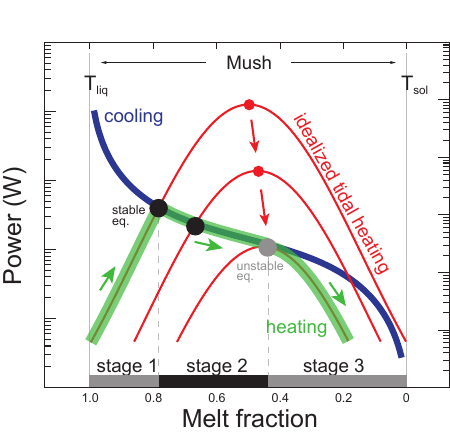}
\end{center}
\hypertarget{fig.-2.-conceptual-three-stage-evolution-of-a-tidally-heated-lunar-magma-ocean.}{%
\subsection*{Fig. 2. Conceptual origin of buffering and collapse in a tidally heated lunar magma ocean.}\label{fig.-2.-conceptual-three-stage-evolution-of-a-tidally-heated-lunar-magma-ocean.}}

Schematic comparison of cooling and tidal-heating power during LMO evolution as a function of melt fraction. The blue curve represents cooling, the green curve represents tidal heating, and the red curves show idealized tidal-heating branches that shift downward as orbital recession proceeds. Black circles mark stable buffered states, whereas the gray circle marks the unstable equilibrium that bounds the high-dissipation branch. Stage 1 corresponds to the earliest, highly molten interval, when the LMO is too liquid to dissipate efficiently and cooling dominates. Stage 2 begins when the system enters the viscoelastic window and becomes trapped in a long-lived tidally buffered state. Stage 3 starts once the stable branch is lost, triggering rapid collapse of tidal heating and renewed cooling.

\clearpage
\begin{center}
\includegraphics[width=0.92\textwidth,height=0.44\textheight,keepaspectratio]{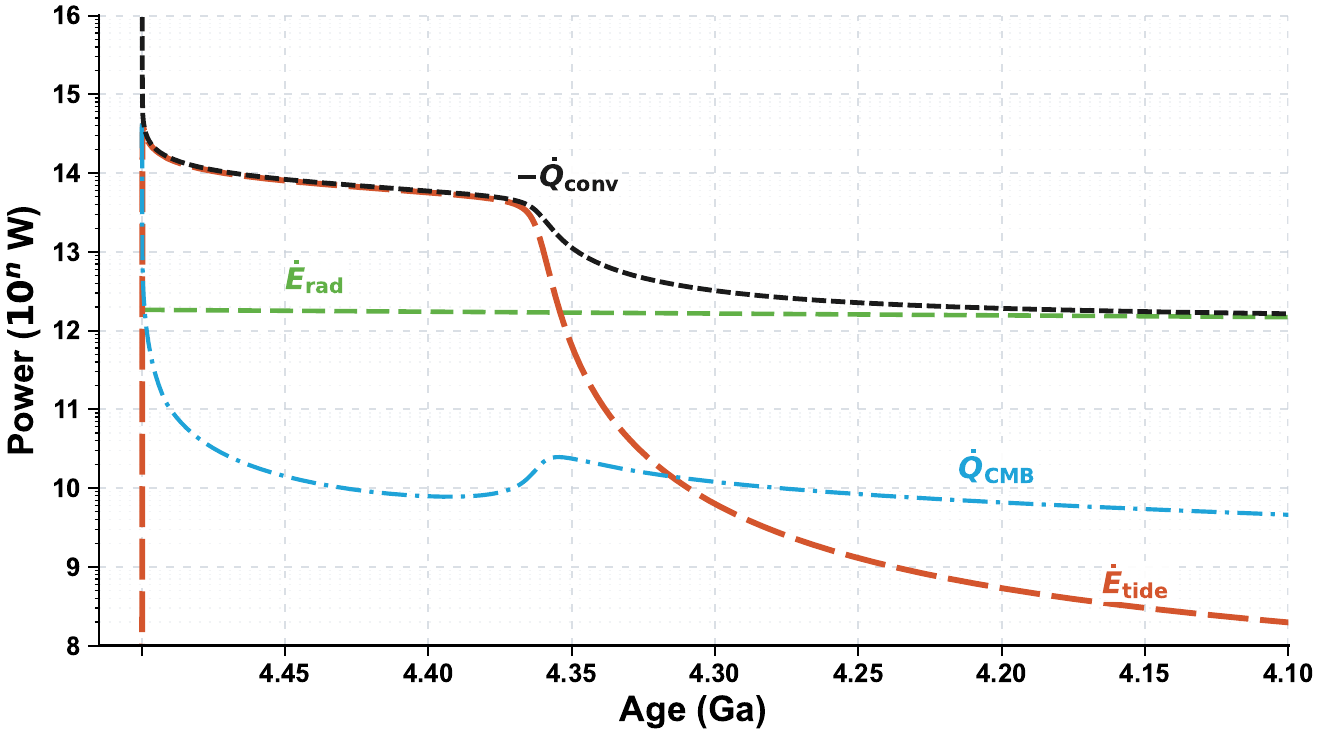}
\end{center}
\hypertarget{fig.-3.-evolution-of-the-major-heating-and-cooling-terms-in-the-baseline-model.}{%
\subsection*{Fig. 3. Evolution of the major heating and cooling terms in the baseline model.}\label{fig.-3.-evolution-of-the-major-heating-and-cooling-terms-in-the-baseline-model.}}

Time evolution of tidal heating $\dot{E}_{\mathrm{tide}}$, radiogenic heating $\dot{E}_{\mathrm{rad}}$, core--mantle boundary heat flux $\dot{Q}_{\mathrm{CMB}}$, and convective heat loss $-\dot{Q}_{\mathrm{conv}}$ from 4.5 to 4.1 Ga in the baseline calculation. Convective cooling dominates immediately after Moon formation. As the LMO enters the viscoelastic window, tidal heating rises sharply and becomes comparable to the dominant cooling terms, producing the long-lived buffered Stage-2 state. Near \textasciitilde4.35 Ga, $\dot{E}_{\mathrm{tide}}$ collapses while cooling terms remain large, initiating renewed net cooling and rapid terminal solidification. The figure also highlights the persistent background secular-cooling sink and radiogenic-heating buffer against which the tidal-heating collapse occurs.

\clearpage
\begin{center}
\includegraphics[width=0.94\textwidth,height=0.46\textheight,keepaspectratio]{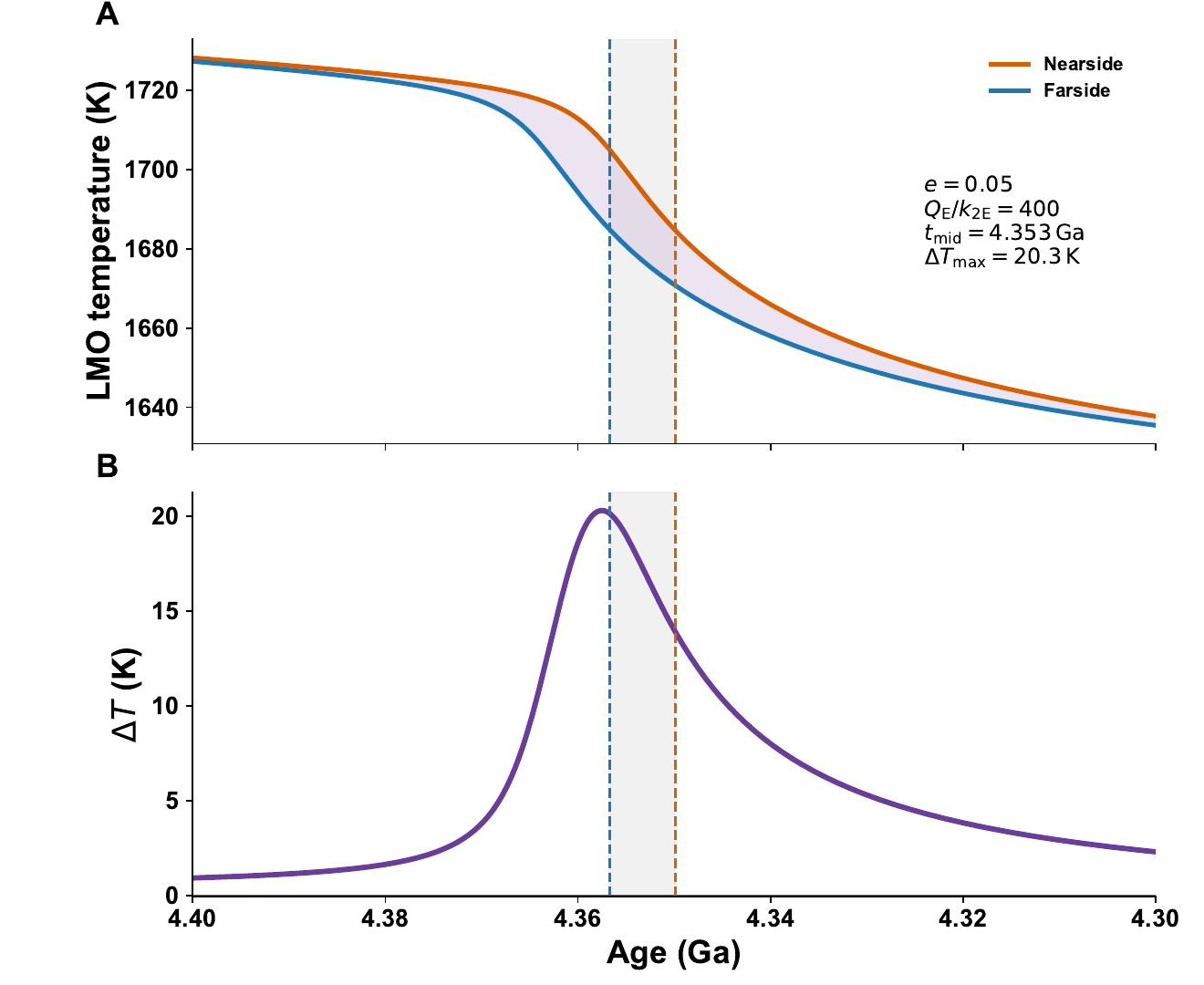}
\end{center}
\hypertarget{fig.-4.-nearside-farside-thermal-asymmetry-during-terminal-lmo-crystallization.}{%
\subsection*{Fig. 4. Baseline nearside-farside temperature divergence during terminal LMO crystallization.}\label{fig.-4.-nearside-farside-thermal-asymmetry-during-terminal-lmo-crystallization.}}

\noindent (A) Modeled LMO temperatures for the nearside and farside in the baseline asymmetric-heating case ($e=0.05$, $Q_{\mathrm{E}}/k_{2\mathrm{E}}=400$). Blue and orange dashed vertical lines mark the modeled farside and nearside thermal-cliff times, respectively, and the gray shaded band denotes the interval between them. The farside reaches the thermal cliff slightly earlier and begins rapid cooling before the nearside, reflecting weaker farside tidal heating and longer nearside buffering. (B) Corresponding nearside-farside temperature difference $\Delta T = T_{\mathrm{near}} - T_{\mathrm{far}}$. The asymmetry remains modest through most of the buffered stage but rises sharply near the cliff, reaching \textasciitilde20 K in the baseline case.

\clearpage
\begin{center}
\includegraphics[width=0.80\textwidth]{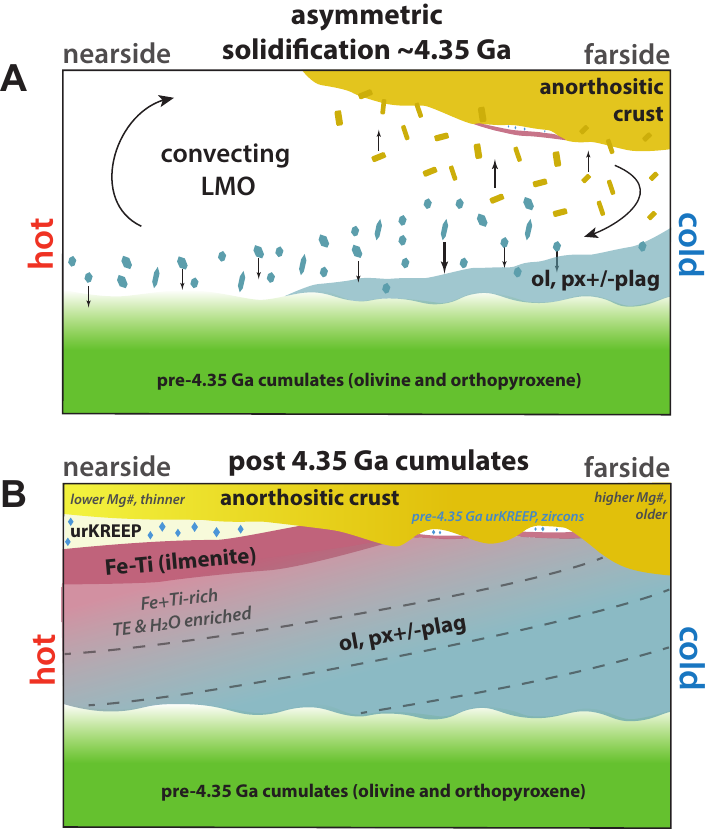}
\end{center}
\hypertarget{fig.-5.-schematic-lunar-consequences-of-asymmetric-late-stage-lmo-crystallization.}{%
\subsection*{Fig. 5. Schematic lunar consequences of asymmetric late-stage LMO crystallization.}\label{fig.-5.-schematic-lunar-consequences-of-asymmetric-late-stage-lmo-crystallization.}}

\noindent (A) State near \textasciitilde4.35 Ga, when the farside has entered rapid cooling while the nearside remains more strongly buffered. Earlier farside cooling promotes earlier plagioclase flotation and thicker anorthositic crust, whereas pre-4.35 Ga mafic cumulates remain broadly similar on both hemispheres. (B) Post-4.35 Ga outcome. Continued asymmetric crystallization concentrates late Fe-Ti-rich and urKREEP-bearing residual products beneath the thinner nearside crust, whereas the farside preserves older, higher-Mg\# anorthosite and thicker crust. Sparse pre-\textasciitilde4.35 Ga zircons are consistent with limited early differentiation, whereas the zircon-age peak near \textasciitilde4.35 Ga reflects bulk terminal solidification; nearside-farside differences in zircon age distributions could test this model. Figure is schematic and not to scale.

\clearpage
\section*{Supplementary Materials}

\renewcommand{\thetable}{S\arabic{table}}
\renewcommand{\theequation}{S\arabic{equation}}
\setcounter{table}{0}
\setcounter{equation}{0}

\noindent The Supplementary Materials comprise two sections. Materials and Methods details the modeling framework, and Supplementary Text presents additional modeling results across a range of parameter values together with further discussions.

\subsection*{Materials and Methods}

The main objective of our model is to estimate the tidal heating within the silicate portion of the Moon following its initially molten state---the so-called Lunar Magma Ocean (LMO)---to shortly after its solidification. Accordingly, our approach follows previous formulations that calculate tidal heating over the full magma ocean life cycle for other bodies. Such models have been developed for Io and exoplanets\textsuperscript{1-6}, as well as for Earth\textsuperscript{7-11}. For the Moon, there have also been studies related to tidal heating targeting the earliest epoch ($\sim$4.5--4.0~Ga), but these works emphasized possible tidal reheating events after solidification of the LMO\textsuperscript{12,13}, rather than dissipation within the magma ocean during its evolution. Hence, to our knowledge, the present study is the first to explicitly simulate tidal heating inside the LMO itself.

We adapt parameterized Io/exoplanet frameworks\textsuperscript{2,3,14} to the Moon and solve for the thermal evolution of the LMO within the context of the Earth--Moon system. The geometry of the model is summarized in Supplementary Fig.~S1. We represent the LMO as a homogeneous layer with uniform temperature and bulk properties. Energy exchange is mediated by parameterized boundary layers at the surface (radiative/advective coupling to space) and at the core--LMO interface (basal heat exchange). All parameter values used in this study are listed in Table~\ref{tab:full_model_parameters_compact}. Given the huge uncertainties in the initial conditions of the LMO and parameters involved, it is reasonable to keep our model simple in order to demonstrate the feasibility of our proposed mechanism in prolonging the existence of the LMO.

\subsection*{1. Energy balance of the LMO}

In our thermal evolution framework, the LMO gains energy from tidal dissipation, radiogenic decay of long-lived nuclides, and possible heat inflow from the core, and it loses energy via convective heat transport eventually lost through its surface. The net energy budget is
\begin{equation}
\dot{E}_{\mathrm{LMO}}
\;=\;
\dot{E}_{\mathrm{tide}}
+ \dot{E}_{\mathrm{rad}}
+ \dot{Q}_{\mathrm{CMB}}
- \dot{Q}_{\mathrm{conv}},
\label{eq:energy_balance}
\end{equation}
where $\dot{E}_{\mathrm{LMO}}$, $\dot{E}_{\mathrm{tide}}$, $\dot{E}_{\mathrm{rad}}$, $\dot{Q}_{\mathrm{CMB}}$, and $\dot{Q}_{\mathrm{conv}}$ are the rates of change of the LMO's internal thermal energy, tidal heating within the LMO, radiogenic heating from long-lived isotopes, the net heat flow into the LMO across the core--mantle boundary (CMB), and the convective heat loss from the LMO to the surface.

The coupled evolution of the LMO mean temperature $T_m$ and the core temperature $T_c$ follows energy conservation within each reservoir. For the LMO,
\begin{equation}
\frac{\mathrm{d} T_m}{\mathrm{d} t}
\;=\;
\frac{\dot{E}_{\mathrm{LMO}}}{C_{\mathrm{LMO,eff}}}
\;=\;
\frac{ \dot{E}_{\mathrm{rad}}
     + \dot{E}_{\mathrm{tide}}
     + \dot{Q}_{\mathrm{CMB}}
     - \dot{Q}_{\mathrm{conv}} }
     { C_{\mathrm{LMO,eff}} } ,
\label{eq:dTm}
\end{equation}
where $C_{\rm LMO,eff}$ is the LMO's effective heat capacity that accounts for latent-heat buffering during crystallization,
\begin{equation}
C_{\mathrm{LMO,eff}}
\;=\;
M_{\mathrm{LMO}}\, c_{\mathrm{LMO}} \,\bigl( 1 + \mathrm{Ste}^{-1} \bigr),
\qquad
\mathrm{Ste}
\;=\;
\frac{c_{\mathrm{LMO}}\,(T_{\mathrm{liq}} - T_{\mathrm{sol}})}{L_{\mathrm{LMO}}} .
\label{eq:Cmeff}
\end{equation}
Here, $M_{\mathrm{LMO}}$ and $c_{\mathrm{LMO}}$ are the LMO mass and specific heat, $L_{\mathrm{LMO}}$ is the latent heat of fusion, and $T_{\mathrm{liq}}$ and $T_{\mathrm{sol}}$ are the liquidus and solidus temperatures, respectively. We adopt $L_{\mathrm{LMO}}=3.2\times10^{5}\ \mathrm{J~kg^{-1}}$\textsuperscript{15}. $\mathrm{Ste}$ is the Stefan number---i.e., the ratio of sensible heat to latent heat---so that the corresponding effective specific heat capacity is $c_{\mathrm{LMO,eff}}=c_{\mathrm{LMO}}+L_{\mathrm{LMO}}/(T_{\mathrm{liq}}-T_{\mathrm{sol}})$.

The core cools through exchange with the LMO across the CMB, which we model as
\begin{equation}
\frac{\mathrm{d} T_c}{\mathrm{d} t}
\;=\;
-\frac{\dot{Q}_{\mathrm{CMB}}}{M_c\,c_c}\,,
\label{eq:dTc}
\end{equation}
where $T_c$ is the core temperature (assumed to be well mixed), $M_c$ the core mass, $c_c$ its specific heat, and $\dot{Q}_{\mathrm{CMB}}$ the net heat flow across the CMB. We adopt the sign convention $\dot{Q}_{\mathrm{CMB}}>0$ for heat flowing from the core into the LMO.

For comparison with the schematic age-cluster cartoons in Fig.~1, we also define an LMO energy-state curve by integrating the same net power budget:
\begin{equation}
E_{\mathrm{LMO}}(t)
\;=\;
E_{\mathrm{LMO}}(t_0)
\;+\;
\int_{t_0}^{t}
\Bigl(
\dot{E}_{\mathrm{tide}}
\;+\;
\dot{E}_{\mathrm{rad}}
\;+\;
\dot{Q}_{\mathrm{CMB}}
\;-\;
\dot{Q}_{\mathrm{conv}}
\Bigr)\,\mathrm{d}t' .
\label{eq:E_LMO_integrated}
\end{equation}
In the present parameterization, $C_{\mathrm{LMO,eff}}$ is treated as constant, so this integrated quantity is equivalently the absolute thermal energy of the LMO reservoir,
\begin{equation}
E_{\mathrm{LMO}}(t) \;=\; C_{\mathrm{LMO,eff}}\,T_m(t),
\label{eq:E_LMO_CeffTm}
\end{equation}
up to the additive constant used to define the zero point in Eq.~\eqref{eq:E_LMO_integrated}. In practice, Eqs.~\eqref{eq:E_LMO_integrated} and \eqref{eq:E_LMO_CeffTm} produce the same time dependence to numerical precision in our baseline calculations. We use this energy evolution---not an instantaneous heating rate---as the green thermal-state guide curve in Fig.~1A. The corresponding time evolution of the major heating and cooling terms is now shown in Fig.~3. For the baseline case ($e=0.05$, $Q_{\mathrm{E}}/k_{2\mathrm{E}}=400$), it shows a very rapid initial energy drop over $<1$~Myr, followed by a long high-energy plateau and a late thermal cliff near $\sim$4.35~Ga. In the artwork of Fig.~1A, the earliest sub-Myr drop is visually compressed and the vertical placement is schematic for clarity, but the timing of the plateau and cliff is taken directly from this computed energy history.

Simulations begin with initial conditions of $T_m(t_0)=3000~\mathrm{K}$ and $T_c(t_0)=3100~\mathrm{K}$. We find that our simulations are relatively insensitive to this choice (both the absolute values and the difference between $T_m$ and $T_c$, see Section~3.1). In the subsections below, we detail the calculations of the radiogenic heating term $\dot{E}_{\mathrm{rad}}$, the CMB exchange $\dot{Q}_{\mathrm{CMB}}$, and the convective transport $\dot{Q}_{\mathrm{conv}}$. Section~2 presents the tidal-heating formulation, $\dot{E}_{\mathrm{tide}}$.

\subsection*{1.1 Radiogenic heating calculation}

Radiogenic heating in the LMO is modeled as the combined contribution of the four dominant long-lived isotopes, $^{232}$Th, $^{238}$U, $^{235}$U, and $^{40}$K. The total radiogenic power of the mantle at time $t$ is
\begin{equation}
\dot{E}_{\mathrm{rad}}(t) \;=\;
M_{\rm LMO} \sum_{i=1}^{4} H_i \, C_i \,
\exp\!\left[ \lambda_i \,\big(t_{\mathrm{f}} - t\big) \right],
\label{eq:Erad}
\end{equation}
\noindent where $H_i$, $C_i$, $\lambda_i$, $t_{\rm f}$, and $t$ are the present-day specific heat production of isotope $i$, the present-day bulk concentration of isotope $i$ in the LMO, the decay constant (where $\lambda_i=\ln 2/t_{1/2,i}$, with half-life $t_{1/2,i}$), total time since lunar formation (i.e., $t_{\rm f}=4.5$~Gyr), and $t$ is the model time, respectively.
Supplementary Fig.~S2 shows the 4.5-billion-year evolution of radiogenic heating rates in Earth and the Moon. Notably, Earth's radiogenic output declines from $\sim 75~\mathrm{TW}$ at $t=0$ to $\sim 18$--$20~\mathrm{TW}$ at $t=4.5~\mathrm{Gyr}$, whereas the Moon's radiogenic output drops from $\sim 1.8~\mathrm{TW}$ to $\sim 0.4$--$0.5~\mathrm{TW}$ (Supplementary Fig.~S2). In comparison, the Moon's absolute level of radiogenic power is modest and its variation over time comparatively small. Over the window emphasized in this study (4.5--4.0~Ga), Earth's radiogenic power decreases from $\sim 75$ to $\sim 60~\mathrm{TW}$, while the Moon's decreases from $\sim 1.8$ to $\sim 1.4~\mathrm{TW}$. Relative to the larger magnitude and variability of the tidal heating we simulate below, radiogenic heating can be viewed as quasi-constant over this interval.

\subsection*{1.2 Convective heat loss and core-mantle boundary heat flux}

Following Renaud and Henning\textsuperscript{2} and Shoji and Kurita\textsuperscript{15}, we treat the LMO as bounded by two thermal boundary layers: an upper boundary layer (UBL) beneath the surface and a lower boundary layer (LBL) coincident with the CMB. The UBL sets the convective heat loss to space and is the dominant cooling pathway, whereas the LBL regulates the upward core heat flux. We determine boundary-layer thicknesses and fluxes using Nusselt number--Rayleigh number (i.e., Nu and Ra) scalings.

\paragraph{Upper boundary layer.}
The UBL thickness $\delta_{\mathrm{u}}$ is related to the Nusselt number via
\begin{equation}
  \mathrm{Nu}_{\mathrm{u}} \;=\; a_c \!\left(\frac{{\rm Ra}_{\mathrm{u}}}{{\rm Ra}_c}\right)^{1/3},
  \qquad
  \delta_{\mathrm{u}} \;=\; \frac{D_m}{2\,\mathrm{Nu}_{\mathrm{u}}},
  \label{eq:Nu_upper}
\end{equation}
with calibration constants $a_c \approx 0.1$ and critical Rayleigh number ${\rm Ra}_c \simeq 1100$.
The UBL Rayleigh number, ${\rm Ra_u}$, uses LMO-average properties and the following temperature drop $(T_m - T_{\mathrm{surf}})$:
\begin{equation}
  Ra_{\mathrm{u}} \;=\;
  \frac{\rho\,g\,\alpha_T\,\bigl(T_{m}-T_{\mathrm{surf}}\bigr)\,D_m^3}{\kappa\,\eta_{\mathrm{u}}},
  \label{eq:Ra_upper}
\end{equation}
where $\rho$ is density, $g$ gravity, $\alpha_T$ thermal expansivity, $D_m$ LMO thickness, $\kappa$ thermal diffusivity, and $\eta_{\mathrm{u}}$ the effective viscosity in the UBL. The conductive flux through the UBL is
\begin{equation}
  F_{\mathrm{conv}} \;=\; \frac{k_m\,\bigl(T_{m}-T_{\mathrm{surf}}\bigr)}{\delta_{\mathrm{u}}} \;=\;
  \mathrm{Nu}_{\mathrm{u}}\,\frac{2k_m\,\bigl(T_{m}-T_{\mathrm{surf}}\bigr)}{D_m},
\end{equation}
and so the total convective power is
\begin{equation}
  \dot{Q}_{\mathrm{conv}} \;=\; 4\pi R_{\mathrm{M}}^{2}\,F_{\mathrm{conv}}.
  \label{eq:Qconv}
\end{equation}
We cap $\mathrm{Nu}_{\mathrm{u}}\le\mathrm{Nu}_{\max}$ and take $\mathrm{Nu}_{\max}=10^{5}$ to avoid unrealistically thin (centimeters to meters) UBL thickness during the early low-viscosity stage.

\paragraph{Lower boundary layer (CMB).}
At the base, the LBL thickness $\delta_{\ell}$ controls the CMB heat exchange. We scale $\delta_{\ell}$ to $\delta_{\mathrm{u}}$ using a viscosity--contrast parameterization\textsuperscript{12,15}:
\begin{equation}
  \delta_{\ell}
  \;=\; \frac{\delta_{\mathrm{u}}}{2}\,
    \bigl[\gamma \,(T_c - T_m)\bigr]^{-1/3}\,
    \exp\!\left(-\frac{\gamma\,(T_c - T_m)}{6}\right),
  \label{eq:delta_lower}
\end{equation}
where $\gamma \simeq 0.011~\mathrm{K}^{-1}$ controls the effective viscosity contrast across the CMB. This yields thinner LBLs for larger $(T_c - T_m)$ while limiting unrealistically large contrasts. The corresponding CMB total heat flux is
\begin{equation}
  \dot{Q}_{\mathrm{CMB}} \;=\; 4\pi R_c^{2}\, \frac{k_m\,\bigl(T_c - T_m\bigr)}{\delta_{\ell}},
  \label{eq:Qcmb}
\end{equation}
with $R_c$ the core radius, $k_m$ the mantle's thermal conductivity, and $\dot{Q}_{\mathrm{CMB}}>0$ denoting heat flowing from the core into the LMO.

\paragraph{Surface boundary condition.}
We approximate the instantaneous surface temperature as a gray-body balance between the globally averaged absorbed insolation and the interior heat release,
\begin{equation}
  T_{\mathrm{surf}}^{4}
  \;=\;
  \frac{1}{\epsilon_v \sigma_B}\,
  \left[
    \frac{(1-A)\,L_\star}{16 \pi a_\star^2}
    \;+\;
    \frac{\dot{Q}_{\mathrm{conv}}}{4\pi R_{\mathrm{M}}^{2}}
  \right],
  \label{eq:Tsurf}
\end{equation}
where $A$ is Bond albedo, $L_\star$ stellar luminosity, $a_\star$ heliocentric distance, $\epsilon_v$ emissivity, $\sigma_B$ the Stefan--Boltzmann constant, and $R_{\rm M}$ is the radius of the Moon.

In summary, Eqs.~(\ref{eq:Nu_upper})--(\ref{eq:Tsurf}) couple LMO convection, CMB exchange, and radiative cooling: the boundary layers set the efficiency of heat transfer, with $\dot{Q}_{\mathrm{conv}}$ governing outward LMO cooling and $\dot{Q}_{\mathrm{CMB}}$ regulating core--LMO exchange.

\subsection*{2. Tidal heating in the LMO}

We compute the LMO tidal heating using a simplified viscoelastic Love-number formalism, in which the imaginary part of the degree-2 Love number, $-\,{\rm Im}\!\left[k_2(\omega)\right]$, quantifies dissipation within the LMO. For a synchronously rotating Moon on an eccentric orbit about Earth, the global tidal power is\textsuperscript{16}
\begin{equation}
  \dot{E}_{\mathrm{tide}}
  \;=\;
  f_{\mathrm{tvf}}\,
  \frac{21}{2}\,
  \bigl[-{\rm Im}\big(k_2(\omega)\big)\bigr]\,
  \frac{G\, M_{\mathrm{E}}^{\,2}\, R_{\mathrm{M}}^{5}}{a_{\mathrm{E}\text{--}\mathrm{M}}^{6}}\,
  n\,e^{2},
  \label{eq:Etide}
\end{equation}
where $G$ is the gravitational constant, $M_{\mathrm{E}}$ the Earth's mass, $a_{\mathrm{E}\text{--}\mathrm{M}}$ the Earth--Moon semi-major axis, $e$ the orbital eccentricity, and $n$ is the mean motion (defined below). For notational brevity, we hereafter write $a$ in place of $a_{\mathrm{E}\text{--}\mathrm{M}}$ unless otherwise noted. The factor $f_{\mathrm{tvf}}$ accounts for the tidal-volume fraction—the fraction of the Moon's volume that participates in dissipation\textsuperscript{2,3}. We adopt an eccentricity of $e=0.05$ as a conservative choice; additional cases are explored in Section~3.2.

\begin{equation}
  n \;\equiv\; \sqrt{\frac{G\,M_{\mathrm{E}}}{a^{3}}}\,.
  \label{eq:n_def}
\end{equation}

As a baseline assumption, we take tidal dissipation to occur throughout the entire silicate portion above an undissipative core of radius $R_c$, which yields
\begin{equation}
  f_{\mathrm{tvf}}
  \;=\;
  \left[\,1 - \left(\frac{R_c}{R_{\mathrm{M}}}\right)^{\!3}\right]\,
  \phi(T_{\mathrm{m}}),
  \label{eq:ftvf_corecase}
\end{equation}
where $\phi(T_{\mathrm{m}})$ is the volume-averaged melt fraction of the LMO, used here as a proxy for the fraction of the silicate shell that remains in a connected magma-ocean/mush state and can participate efficiently in tidal dissipation. In practice, the fraction of tidal power actually generated is likely smaller than the full-silicate-portion value; we revisit this point in Section~3.3.

\subsection*{2.1 Imaginary part of the Love number}
\label{sec:Imk2}

The dissipation is governed by the negative imaginary part of the degree-2 Love number, $-\,{\rm Im}(k_2)$, which encapsulates the frequency-dependent lag between forcing and response. Following Efroimsky\textsuperscript{5},
\begin{equation}
  -\,{\rm Im}(k_2) \;=\; -\,\frac{3}{2}\,
  \frac{J_U\,\bar{\mu}\,{\rm Im}(\tilde{J})}
       {\bigl[{\rm Im}(\tilde{J})\bigr]^2 + \bigl[{\rm Re}(\tilde{J}) + J_U \bar{\mu}\bigr]^2 } ,
  \label{eq:imk2}
\end{equation}
where $\tilde{J}(\omega)$ is the complex compliance of the viscoelastic body, $J_U = 1/\mu_U$ is the unrelaxed (instantaneous) compliance set by the infinite-frequency shear modulus $\mu_U$, and
\begin{equation}
  \bar{\mu} \;=\; \frac{19\,\mu_U}{2\,\rho\,g\,R_{\mathrm{M}}}
  \label{eq:mu_tilde}
\end{equation}
is the (dimensionless) effective rigidity where $\rho$ is bulk density and $g$ surface gravity. Here, ${\rm Re}(\tilde{J})$ and ${\rm Im}(\tilde{J})$ denote the real and imaginary parts of the complex compliance $\tilde{J}$, respectively. Physically, the real part describes elastic storage, whereas the imaginary part quantifies dissipation.

\subsection*{2.2 Complex compliance: Maxwell model for the LMO}

We adopt a Maxwell rheology for the LMO, with complex compliance
\begin{equation}
  \tilde{J}_M(\omega) \;=\; J_U \;+\; \frac{1}{i\,\omega\,\eta_S}
  \;=\; J_U \;-\; i\,\frac{1}{\omega\,\eta_S},
  \label{eq:J_Maxwell}
\end{equation}
where $\eta_S$ is the shear viscosity and $\omega$ is the principal tidal forcing frequency. Thus ${\rm Re}(\tilde{J}_M)=J_U$ and ${\rm Im}(\tilde{J}_M)=-1/(\omega\eta_S)$.

Substituting Eq.~\eqref{eq:J_Maxwell} into Eq.~\eqref{eq:imk2} yields a closed-form expression for the $k_2$ Love number:
\begin{equation}
  -\,{\rm Im}(k_2)_{\!M}
  \;=\;
  \frac{3}{2}\;
  \frac{J_U\,\bar{\mu}\,(\omega\,\eta_S)}
       {1 + \big[J_U(1+\bar{\mu})\,(\omega\,\eta_S)\big]^2 } .
  \label{eq:imk2_maxwell}
\end{equation}
Equations~\eqref{eq:Etide} and \eqref{eq:imk2_maxwell} couple the orbital state $(a,e,n)$ with the LMO rheology $(\mu_U,\eta_S,\rho,g)$ to give the time-dependent tidal power.

\subsection*{2.3 Transitional rheology parameterizations}

\paragraph{(i) Melt fraction.} Following Zahnle et al.\textsuperscript{7}, melt fraction is expressed as
\begin{equation}
\phi(T)=
\begin{cases}
0, & T \le T_{\rm sol},\\[4pt]
\dfrac{T-T_{\rm sol}}{T_{\rm liq}-T_{\rm sol}}, & T_{\rm sol} < T < T_{\rm liq},\\[10pt]
1, & T \ge T_{\rm liq},
\end{cases}
\label{eq:phi_def}
\end{equation}

\paragraph{(ii) Shear modulus.}
Following Henning et al.\textsuperscript{3}, we adopt a three-piece liquidus-plateau form
\begin{equation}
\mu_U(T)=
\begin{cases}
\mu_0, & T \le T_{\rm sol},\\[3pt]
\mu_0\,\exp\!\left[\dfrac{E_a}{R}\left(\dfrac{1}{T}-\dfrac{1}{T_{\rm sol}}\right)\right], & T_{\rm sol}<T<T_{\rm liq},\\[8pt]
\mu_0\,\exp\!\left[\dfrac{E_a}{R}\left(\dfrac{1}{T_{\rm liq}}-\dfrac{1}{T_{\rm sol}}\right)\right], & T \ge T_{\rm liq},
\end{cases}
\label{eq:mu_hos}
\end{equation}
where $\mu_0$ is the solidus reference modulus at $T_{\rm sol}=1600\ \mathrm{K}$, 
$E_a$ is the activation energy, and $R$ is the universal gas constant.
This form is equivalent to $\mu_0\exp(E_a/RT-C_0)$ with $C_0=E_a/(R\,T_{\rm sol})$.

\paragraph{(iii) Viscosity.}
As described in the main text, the essence of the enhanced tidal heating is that the dimensionless parameter $\chi \equiv \omega\tau$ is near unity ($\chi \approx 1$), where $\omega$ is the tidal forcing frequency and $\tau$ is the viscoelastic relaxation time. For lunar forcing frequencies this places the peak tidal-heating viscosity of the LMO at $\eta_S \sim 10^{12}$--$10^{16}\,\mathrm{Pa\,s}$. Accordingly, we adopt a temperature-segmented parameterization with three anchor values:
$\eta_{S,\mathrm{sol}}^{\mathrm{ref}}$, $\eta_{S,\mathrm{cri}}^{\mathrm{ref}}$, and $\eta_{S,\mathrm{liq}}^{\mathrm{ref}}$.
These represent the shear viscosities at $T_{\mathrm{sol}}$, $T_{\mathrm{cri}}$, and $T_{\mathrm{liq}}$, respectively; with default values used in our runs
\(\eta_{S,\rm sol}^{\rm ref}=10^{19}\ {\rm Pa\,s}\),
\(\eta_{S,\rm cri}^{\rm ref}=3.5\times10^{12}\ {\rm Pa\,s}\),
\(\eta_{S,\rm liq}^{\rm ref}=10^{-3}\ {\rm Pa\,s}\). Thus, by tuning the viscosity near $T_{\mathrm{cri}}$, we locate the tidal heating maximum within the transitional window at $\phi_{\mathrm{crit}} \approx 0.4$ (i.e., where $\chi \simeq 1$). Between these anchors, we parameterize the viscosity with an Arrhenius backbone augmented by a melt-weakening mechanism. Let $A_\phi$ be the weakening coefficient and set $\phi_s=0$, $\phi_{\mathrm{crit}}=(T_{\rm cri}-T_{\rm sol})/(T_{\rm liq}-T_{\rm sol})$,
$\phi_l=1$.

The shear viscosity is prescribed as
\begin{equation}
\eta_S(T)=
\begin{cases}
\eta_{S,\rm sol}^{\rm ref}\,
\exp\!\Big[\dfrac{E_{a}}{R}\Big(\dfrac{1}{T}-\dfrac{1}{T_{\rm sol}}\Big)\Big],
& T\le T_{\rm sol},\\[10pt]
\exp\!\big(a_1 + b_1\,T^{-1}\big)\;\exp\!\big(-A_\phi\,\phi(T)\big),
& T_{\rm sol}<T\le T_{\rm cri},\\[6pt]
\exp\!\big(a_2 + b_2\,T^{-1}\big)\;\exp\!\big(-A_\phi\,\phi(T)\big),
& T_{\rm cri}<T\le T_{\rm liq},\\[6pt]
\exp\!\big(a_2 + b_2\,T_{\rm liq}^{-1}\big)\;\exp\!\big(-A_\phi\,\phi_l\big),
& T> T_{\rm liq},
\end{cases}
\label{eq:eta_case5}
\end{equation}
where $R$ is the gas constant and $E_{a}$ is the activation energy for the low-temperature solid branch (no $\phi$-weakening). The coefficients $a_1,b_1$ and $a_2,b_2$ are determined by the two-point constraints in each segment. This formulation ensures continuity at $T_{\rm sol}$, $T_{\rm cri}$, and $T_{\rm liq}$, keeps the high-temperature branch constant at the (already weakened) value at $T_{\rm liq}$, and reduces to a pure Arrhenius law below the solidus. In the numerics we cap $\eta_S$ to $[\eta_{\min},\eta_{\max}]$ with
\(\eta_{\min}=10^{-6}\) and \(\eta_{\max}=10^{27}\ {\rm Pa\,s}\).

\paragraph{Peak viscosity.}
The dissipation efficiency is set by $-{\rm Im}[k_2(\omega)]$, whose Maxwell form is already given in Eq.~\eqref{eq:imk2_maxwell}. Let $x\equiv\omega\eta_S$. Maximizing $-{\rm Im}[k_2]$ with respect to $x$ yields
\begin{equation}
x_{\rm peak}=\frac{1}{J_U\,(1+\bar{\mu})}\,,
\qquad
\omega\,\tau_{\rm eff}=1,
\qquad
\tau_{\rm eff}=\frac{(1+\bar{\mu})\,\eta_S}{\mu_U},
\label{eq:xpeak_taueff}
\end{equation}
and the corresponding peak value $-{\rm Im}[k_2]_{\rm peak}= \tfrac{3}{4}\,\bar{\mu}/(1+\bar{\mu})$.
Hence the viscosity at which dissipation peaks is
\begin{equation}
\eta_{\rm peak}=\frac{\mu_U}{(1+\bar{\mu})\,\omega}.
\label{eq:eta_peak_general}
\end{equation}

In the rigid limit ($\bar{\mu}\gg1$) and using
$\bar{\mu}=19\mu_U/(2\rho g R_{\mathrm{M}})$, Eq.~\eqref{eq:eta_peak_general}
reduces to the simple scaling
\begin{equation}
\eta_{\rm peak}\approx \frac{2\,\rho\,g\,R_{\mathrm{M}}}{19\,\omega},
\label{eq:eta_peak_rigid}
\end{equation}
where all material and geometric parameters are as listed in Table~S1.
For a synchronously rotating Moon the principal forcing frequency is the mean
motion, $\omega\simeq n=\sqrt{G M_{\mathrm{E}}/a^{3}}$ [Eq.~\eqref{eq:n_def}],
so that
\begin{equation}
\eta_{\rm peak}(a)\approx
\frac{2\,\rho\,g\,R_{\mathrm{M}}}{19}
\left(\frac{a^{3}}{G M_{\mathrm{E}}}\right)^{1/2}
\propto a^{3/2}.
\label{eq:eta_peak_of_a}
\end{equation}

With the lunar parameters in Table~S1, $\eta_{\rm peak}(a)$ over
$a=3$--$60\,R_{\mathrm{E}}$ lies in the transitional window targeted by our
parameterization (order $10^{12}$--$10^{16}\ \mathrm{Pa\,s}$), and we illustrate
this in Supplementary Fig.~S3 with a conservative $\times10$ envelope reflecting uncertainty
in frequency content and rheology.

\subsection*{2.4 Earth--Moon distance evolution}

The secular evolution of the lunar semi-major axis $a(t)$ due to tides raised on the Earth is modeled with the constant $k_{2,\mathrm{E}}/Q_{\mathrm{E}}$ formulation following Nimmo et al.\textsuperscript{17} and Murray and Dermott\textsuperscript{18}. The orbital recession rate is
\begin{equation}
  \frac{da}{dt}
  \;=\;
  3 \left(\frac{k_{2,\mathrm{E}}}{Q_\mathrm{E}}\right)
    \left(\frac{M_\mathrm{M}}{M_\mathrm{E}}\right)
    \left(\frac{R_\mathrm{E}}{a}\right)^{5}
    n\,a,
  \label{eq:da_dt}
\end{equation}
where $G$ is the gravitational constant, $M_{\rm E}$, $R_{\rm E}$, $k_{2,\rm E}$, and $Q_{\rm E}$ are the Earth's mass, radius, degree-2 tidal Love number, and tidal quality factor. $M_{\rm M}$ is the lunar mass and $n$ is the mean motion defined in Eq.~\eqref{eq:n_def}. The initial semi-major axis is set to $a_0 = 3.0\,R_\mathrm{E}$, slightly beyond the Roche limit. In our baseline case we adopt constant values $k_{2,\mathrm{E}} = 0.3$ and $Q_\mathrm{E} = 120$, so that $Q_\mathrm{E}/k_{2,\mathrm{E}} = 400$. For the very early epoch considered here, Earth's effective $k_{2,\mathrm{E}}$ and $Q_\mathrm{E}$ are uncertain and could differ substantially from present-day values; a hotter, more deformable Earth could exhibit higher $k_{2,\mathrm{E}}$ and higher $Q_\mathrm{E}$ (i.e., less dissipation per cycle). Using this constant ratio provides a simple, conservative baseline that yields an Earth--Moon distance history consistent with previous work\textsuperscript{6}, anchoring $a \approx 35\,R_\mathrm{E}$ at $4.0~\mathrm{Ga}$ without invoking a closer initial configuration (Supplementary Fig.~S3C). Additional Earth--Moon distance evolutions are examined in Section~3.2.

\subsection*{2.5 Nearside--farside tidal heating calculation}

In our baseline symmetric model, the tidally dissipated power
$\dot{E}_{\mathrm{tide}}$ depends on the center-to-center Earth--Moon
distance $a_{\mathrm{E}\text{--}\mathrm{M}}$ and the forcing frequency
through Eqs.~\eqref{eq:n_def} and \eqref{eq:Etide}. For convenience we
define a ``baseline'' tidal power at the lunar center,
\begin{equation}
\dot{E}_{\mathrm{tide,0}}(a_{\mathrm{E}\text{--}\mathrm{M}},T_m)
  \;\equiv\;
  \dot{E}_{\mathrm{tide}}\!\left(
    a=a_{\mathrm{E}\text{--}\mathrm{M}},\,
    \omega=n(a_{\mathrm{E}\text{--}\mathrm{M}});\,
    T_m
  \right),
\label{eq:Etide0_def}
\end{equation}
where the rheology (and thus $-{\rm Im}[k_2]$) is evaluated using the local
mantle temperature $T_m$, while the forcing frequency and Love number are
always tied to the same center-to-center distance
$a_{\mathrm{E}\text{--}\mathrm{M}}$.

The dominant degree-2 tide in this baseline model is symmetric between the
nearside and farside. The hemispheric asymmetry introduced below should
therefore be understood as a simple higher-order finite-distance correction to
the tide-raising potential, which becomes more relevant when the early Moon is
close to Earth\textsuperscript{19}. If the lunar interior had already
developed lateral heterogeneity, then degree-2 forcing could also couple into
higher-degree tidal responses; recent degree-3 Love-number results suggest that
this may matter for the present Moon\textsuperscript{20}. We do not model
that structural coupling here, and instead isolate only the direct finite-size
contribution to the near--far heating contrast.

To estimate the near--far asymmetry, we treat the nearside and farside as small
radial offsets $\pm R_{\mathrm{M}}$ from the lunar center along the Earth--Moon
line. The Earth's gravitational acceleration at radius $r$ is
\begin{equation}
g(r) \;=\; \frac{G M_{\rm E}}{r^2},
\end{equation}
and we write $r = a_{\mathrm{E}\text{--}\mathrm{M}}(1 + \delta)$ with
$|\delta|\ll 1$. A Taylor expansion of $g(r)$ about
$r = a_{\mathrm{E}\text{--}\mathrm{M}}$ gives
\begin{equation}
g(r)
  \;=\;
  \frac{G M_{\rm E}}{a_{\mathrm{E}\text{--}\mathrm{M}}^{2}}\,(1+\delta)^{-2}
  \;\approx\;
  g_0\,(1 - 2\delta + 3\delta^2),
\qquad
g_0 \;\equiv\; g(a_{\mathrm{E}\text{--}\mathrm{M}})
             \;=\; \frac{G M_{\rm E}}{a_{\mathrm{E}\text{--}\mathrm{M}}^{2}},
\end{equation}
where $g_0$ is the Earth's gravitational acceleration at the lunar center.
Defining the dimensionless finite-size parameter
\begin{equation}
\varepsilon_R \;\equiv\; \frac{R_{\mathrm{M}}}{a_{\mathrm{E}\text{--}\mathrm{M}}},
\end{equation}
the nearside and farside correspond to
$\delta_{\mathrm{near}}=-\varepsilon_R$ and
$\delta_{\mathrm{far}}=+\varepsilon_R$, respectively. We note importantly that $\varepsilon_R$ is time dependent and decreases significantly as $a_{\mathrm{E}\text{--}\mathrm{M}}$ increases. At present, higher-order terms in $\delta$ are negligible; earlier in Earth--Moon evolution, they are less so. The local tidal
acceleration along the Earth--Moon line, which is proportional to the
gradient of the tidal potential, is the difference between $g(r)$
and $g_0$, so that, to second order in $\varepsilon_R$,
\begin{align}
\Delta g_{\mathrm{near}}
  &\equiv g(a_{\mathrm{E}\text{--}\mathrm{M}}-R_{\mathrm{M}})-g_0
   \;\approx\;
   g_0\bigl(2\varepsilon_R+3\varepsilon_R^2\bigr),\\
\Delta g_{\mathrm{far}}
  &\equiv g(a_{\mathrm{E}\text{--}\mathrm{M}}+R_{\mathrm{M}})-g_0
   \;\approx\;
   g_0\bigl(-2\varepsilon_R+3\varepsilon_R^2\bigr).
\end{align}
We factor out the leading-order (point-mass) tidal amplitude,
\begin{equation}
\Delta g_0 \;\equiv\; 2\varepsilon_R\,g_0,
\end{equation}
and obtain, to first order in $\varepsilon_R$,
\begin{equation}
\frac{\Delta g_{\mathrm{near}}}{\Delta g_0}
  \;\approx\; 1 + \tfrac{3}{2}\varepsilon_R,
\qquad
\frac{\Delta g_{\mathrm{far}}}{\Delta g_0}
  \;\approx\; 1 - \tfrac{3}{2}\varepsilon_R.
\end{equation}

Assuming that tidal strain amplitude scales with $\Delta g$ and that
dissipated power scales with the square of the strain, the corresponding
hemispheric correction factors for the tidally dissipated power are
\begin{equation}
C_{\mathrm{near}}
  \;=\;
  \bigl(1 + \tfrac{3}{2}\varepsilon_R\bigr)^2,
\qquad
C_{\mathrm{far}}
  \;=\;
  \bigl(1 - \tfrac{3}{2}\varepsilon_R\bigr)^2,
\label{eq:C_near_far_simple}
\end{equation}
which both reduce to unity in the point-mass limit
$R_{\mathrm{M}}/a_{\mathrm{E}\text{--}\mathrm{M}}\to0$.

At each time step we therefore compute nearside and farside tidal powers
simply as
\begin{align}
\dot{E}_{\mathrm{tide,near}}(t)
  &= C_{\mathrm{near}}\!\bigl(a_{\mathrm{E}\text{--}\mathrm{M}}(t)\bigr)\,
     \dot{E}_{\mathrm{tide,0}}\bigl(a_{\mathrm{E}\text{--}\mathrm{M}}(t),
                                    T_{m,\mathrm{near}}(t)\bigr),\\
\dot{E}_{\mathrm{tide,far}}(t)
  &= C_{\mathrm{far}}\!\bigl(a_{\mathrm{E}\text{--}\mathrm{M}}(t)\bigr)\,
     \dot{E}_{\mathrm{tide,0}}\bigl(a_{\mathrm{E}\text{--}\mathrm{M}}(t),
                                    T_{m,\mathrm{far}}(t)\bigr),
\end{align}
and evolve two otherwise identical zero-dimensional thermal columns for
the nearside and farside using these hemispheric tidal powers. The
center-to-center distance $a_{\mathrm{E}\text{--}\mathrm{M}}(t)$ remains
governed by the same Earth-dominated tidal torque as in the symmetric
model. The resulting near--far differences in temperature and cliff age
are shown in Fig.~4 and Supplementary Fig.~S6.

\subsection*{Supplementary Text}

\subsection*{3. Parameter-space exploration}

\subsection*{3.1 Sensitivity to initial temperatures}

We assess the sensitivity of the tidal heating trajectory to the initial LMO and core temperatures $(T_{m0},T_{c0})$ while holding orbital configuration fixed (i.e., $e=0.05,\; Q_{\mathrm{E}}/k_{2\mathrm{E}}=400$ for all time) and the initial semi-major axis $a_0=3.0\,R_{\mathrm{E}}$, matching the example case in the main text. As shown in Supplementary Fig.~S4, different $(T_{m0},T_{c0})$ pairs affect only the very early ($\gtrsim 4.499\,\mathrm{Ga}$) transient period---specifically, the amplitude of the initial $\dot E_{\mathrm{tide}}$ spike and the precise entry time into the transitional rheology window (where the global melt fraction $\approx 0.7$). After this brief start-up, the $\dot E_{\mathrm{tide}}(t)$ trajectories rapidly converge and the resulting cliff ages---defined by the solutions of $\dot E_{\mathrm{tide}}=\dot E_{\mathrm{rad}}$---are virtually indistinguishable across all tested initial temperatures. Panels B--C of Supplementary Fig.~S4 further show that varying the initial core--mantle temperature offset, $\Delta T_{c\!-\!m}$ (including $T_{c0}=T_{m0}$ and $T_{c0}=T_{m0}+50$--$100\,\mathrm{K}$), does not appreciably alter the post-transient evolution or the inferred cliff ages; hence, the model is insensitive to both $(T_{m0},T_{c0})$ and $\Delta T_{c\!-\!m}$. 

\subsection*{3.2 Eccentricity and Earth--Moon distance evolution}

Lunar orbital parameters during the LMO epoch are uncertain and lack direct material evidence. As summarized in Eq.~\eqref{eq:Etide}, the dominant orbital controls in our tidal calculation are the semi-major axis $a$ and the eccentricity $e$; to first order, tidal heating scales as $a^{-6}$ and $e^{2}$. Today, $a \approx 60\,R_{\mathrm{E}}$ and $e \approx 0.055$. At earlier times the Moon likely orbited closer to Earth with a larger $e$, implying stronger tidal dissipation than later in its history. For simplicity, we prescribe a time-varying semi-major axis $a(t)$ and a constant eccentricity over the interval of interest.

Supplementary Fig.~S5 provides a detailed overview of tidal-heating trajectories. Across variations in eccentricity $e$ or the ratio $Q_{\mathrm{E}}/k_{2\mathrm{E}}$, all curves exhibit a robust high-plateau--sharp-cliff pattern. Increasing $e$ or $Q_{\mathrm{E}}/k_{2\mathrm{E}}$ (i.e., decreasing the Earth--Moon distance and thus strengthening the effective tidal forcing) sustains higher tidal power for a longer period of time and yields systematically younger cliff ages. Conversely, smaller values lead to an earlier cliff closer to $4.5\,\mathrm{Ga}$. 

The evolution of $a(t)$ is governed primarily by tides raised by the Moon on Earth\textsuperscript{18}. The recession rate therefore depends on Earth’s tidal response, which we express through the ratio $Q_{\mathrm{E}}/k_{2\mathrm{E}}$; it shows a strong inverse dependence on $a(t)$ (see Eq.~\eqref{eq:da_dt}). In the example run of Fig.~1, we adopt $Q_{\mathrm{E}}/k_{2\mathrm{E}}=400$, yielding $a \approx 29.4\,R_{\mathrm{E}}$ at $4.35~\mathrm{Ga}$. This matches previously published reconstructions (see Supplementary Fig.~S3C). This choice is also consistent with inclination constraints: strong LMO tidal dissipation would remove orbital inclination. The present lunar inclination of about $5.16^{\circ}$ implies that LMO damping could not have persisted while the Moon lingered near $a \approx 20\,R_{\mathrm{E}}$, where damping at low obliquity is most efficient\textsuperscript{21}. In addition, obliquity can become large near the Cassini-state transition at $a \approx 30\,R_{\mathrm{E}}$\textsuperscript{21-23}; large obliquity greatly enhances obliquity tides, so if the LMO were still dissipative, the inclination would be efficiently damped and a primordial value erased. To avoid such overprinting, the LMO must have largely solidified before the system spent substantial time near that threshold. Accordingly, our adopted $Q_{\mathrm{E}}/k_{2\mathrm{E}}=400$ should be viewed as a practical lower bound that preserves the present inclination under a highly dissipative LMO. It is also worth noting that Earth’s tidal dissipation in the early epoch may have undergone its own transitional rheology, implying strong time variability in $Q_{\mathrm{E}}$---from low values during highly dissipative states to potentially extremely high values (on the order of $10^{12}$--$10^{16}$) when dissipation is weak. Pinning down when Earth’s magma ocean solidified, and the corresponding tidal parameters and their impact on the Moon, is important future work but beyond our present scope.

Lunar eccentricity is set by a balance between tidal damping and resonant or secular excitation like evection. Published solutions allow much larger ancient values than our baseline: for synchronous rotation at $a \approx 22.9\,R_{\mathrm{E}}$, $e \approx 0.49$ has been inferred, and evection can pump $e$ to $\sim 0.5$ at $a \approx 4.6\,R_{\mathrm{E}}$\textsuperscript{24}. By comparison, we adopt $e=0.05$, which lies well below these ranges and is therefore very conservative for the interval considered.

Hence, our baseline case adopts a relatively large Earth--Moon distance and a small eccentricity, so the enhanced tidal heating is not driven primarily by orbital forcing but by the LMO’s rheological state. 

\subsection*{3.3 Effective volume fraction}

Our baseline simulation assumed that the entire LMO participates in the tidal response (effective fraction $f\cdot\phi(T_{\mathrm{m}})$ with $f=1$). To test partial participation, we scale tidal dissipation by an effective participating fraction $f\in(0,1]$ and, for the baseline orbital configuration (i.e., where $e=0.05,\; Q_{\mathrm{E}}/k_{2\mathrm{E}}=400$), we reduce $f$ systematically to $0.8,\,0.6,\,0.4,\,0.2$, testing the entire evolution each time.
The result is a uniformly weaker $\dot E_{\mathrm{tide}}$, a shortened high-energy plateau, and systematically older cliff ages (defined by the crossing $\dot E_{\mathrm{tide}}=\dot E_{\mathrm{rad}}$). However, increasing the eccentricity to $e=0.10$---a plausible value for early lunar orbits\textsuperscript{6}---compensates for small $f$, yielding cliff ages near $\sim 4.35\,\mathrm{Ga}$ or younger. Thus, allowing only a fraction of the LMO to enter the transitional rheology regime still preserves the robust high-plateau--sharp-cliff pattern. Importantly, the $f=1$ baseline value should not be viewed as an overestimate, given our conservative baseline value for $e$ and the effective forcing ratio $Q_{\mathrm{E}}/k_{2\mathrm{E}}$.

\subsection*{3.4 Age and temperature asymmetry of the LMO}

Our model predicts that tidal dissipation differs between the lunar nearside and farside and, consequently, that the terminal stages of LMO crystallization may develop asymmetries in both temperature and crystallization age (approximated here by the ``cliff age''). We chart these asymmetries across the $(e,\;Q_{\mathrm{E}}/k_{2\mathrm{E}})$ space by quantifying the peak near--far temperature contrast $\max|\Delta T_{\mathrm{m}}|$ and the cliff-age difference $\Delta t_{\mathrm{cliff}}\!\equiv\! t_{\mathrm{cliff}}^{\mathrm{near}}-t_{\mathrm{cliff}}^{\mathrm{far}}$ (Supplementary Fig.~S6). For $e\!\approx\!0.10$ and $Q_{\mathrm{E}}/k_{2\mathrm{E}}\!\approx\!800$, the model yields a temperature contrast exceeding $85\ \mathrm{K}$ and a cliff-age offset of $\sim80\ \mathrm{Myr}$. Our setup remains simplified; future work should incorporate more realistic geometry, rheology, and orbit--thermal coupling.

\subsection*{3.5 Possible feedback of tidal dissipation on lunar eccentricity}

In the calculations presented here, the eccentricity $e$ is prescribed in order to isolate the thermal consequences of tidal heating. In reality, however, the same dissipation that heats the LMO should also react back on the orbit. A more complete treatment would therefore evolve the eccentricity together with the semi-major axis, the thermal state of the Moon, and the tidal/spin state of the Earth. At leading order, the eccentricity budget may be written schematically as
\begin{equation}
  \frac{de}{dt}
  \;=\;
  \left(\frac{de}{dt}\right)_{\mathrm{E}}
  \;+\;
  \left(\frac{de}{dt}\right)_{\mathrm{M}}
  \;+\;
  \left(\frac{de}{dt}\right)_{\mathrm{res}},
  \label{eq:de_dt_total}
\end{equation}
where the three terms represent, respectively, tides raised on the Earth by the Moon, tides raised on the Moon by the Earth, and resonant or secular forcing (e.g., evection) that is not included here\textsuperscript{17,18,21}.

The Earth term can either excite or damp eccentricity depending on the terrestrial spin rate. In a commonly used small-$e$ constant-$Q$ formulation\textsuperscript{18,21},
\begin{equation}
  \left(\frac{de}{dt}\right)_{\mathrm{E}}
  \;\approx\;
  \frac{57}{8}
  \left(\frac{k_{2,\mathrm{E}}}{Q_{\mathrm{E}}}\right)
  \left(\frac{M_{\mathrm{M}}}{M_{\mathrm{E}}}\right)
  \left(\frac{R_{\mathrm{E}}}{a}\right)^5
  n e
  \left(
    \frac{11}{18}\frac{\Omega_{\mathrm{E}}}{n} - 1
  \right),
  \label{eq:de_dt_earth}
\end{equation}
where $\Omega_{\mathrm{E}}$ is the Earth's spin angular velocity. Thus, when the early Earth rotates sufficiently rapidly ($\Omega_{\mathrm{E}} > 18n/11$), tides in Earth can pump $e$; after terrestrial spin slows, the same term becomes damping. The Earth contribution is therefore sensitive not only to $Q_{\mathrm{E}}/k_{2,\mathrm{E}}$, but also to the poorly constrained spin and rheological evolution of the early Earth.
It also depends implicitly on the orbital distance itself, because both $(R_{\mathrm{E}}/a)^5$ and the mean motion $n=\sqrt{GM_{\mathrm{E}}/a^3}$ vary as the Moon recedes. Thus, even before the explicit Moon term is considered, the eccentricity history is already coupled to the uncertain evolution of $a(t)$.

The lunar term is more naturally damping for a synchronously rotating Moon. In a classical constant-$Q$ representation it may be written as
\begin{equation}
  \left(\frac{de}{dt}\right)_{\mathrm{M}}
  \;\approx\;
  -\frac{21}{2}
  \left(\frac{k_{2,\mathrm{M}}}{Q_{\mathrm{M}}}\right)
  \left(\frac{M_{\mathrm{E}}}{M_{\mathrm{M}}}\right)
  \left(\frac{R_{\mathrm{M}}}{a}\right)^5
  n e,
  \label{eq:de_dt_moon_classical}
\end{equation}
where $k_{2,\mathrm{M}}/Q_{\mathrm{M}}$ should be understood as an effective lunar tidal response. In the present framework, because the LMO dissipation is computed explicitly through $\dot E_{\mathrm{tide}}$, an equivalent and often more convenient form is obtained by relating the dissipated power to the small-$e$ epicyclic orbital energy,
\begin{equation}
  E_{e}
  \;\approx\;
  \frac{1}{2}\,\frac{G M_{\mathrm{E}} M_{\mathrm{M}}}{a}\,e^2,
  \qquad
  \left(\frac{de}{dt}\right)_{\mathrm{M}}
  \;\approx\;
  -\,\frac{(1-e^2)\,a}{G M_{\mathrm{E}} M_{\mathrm{M}}\,e}\,
  \dot E_{\mathrm{tide}} .
  \label{eq:de_dt_moon_energy}
\end{equation}
An additional complication is that the effective lunar tidal response is not fixed in the present framework. In a classical formulation, Eq.~\eqref{eq:de_dt_moon_classical} would be evaluated with a prescribed $k_{2,\mathrm{M}}/Q_{\mathrm{M}}$, but here that quantity should be understood as evolving with the thermal and rheological state of the LMO. Because eccentricity affects the heating rate, and the heating rate in turn modifies temperature, melt fraction, and viscosity, changes in $e$ can indirectly alter the Moon's effective $k_{2,\mathrm{M}}/Q_{\mathrm{M}}$ as well. Thus the lunar damping term is doubly nonlinear: it depends explicitly on $e$ and $a$, and implicitly on the state-dependent rheology that those same orbital variables help regulate.
This expression makes the physical feedback particularly transparent: during the high-dissipation Stage~2, large $\dot E_{\mathrm{tide}}$ should also drive strong eccentricity damping, whereas after the thermal cliff the collapse of $\dot E_{\mathrm{tide}}$ sharply weakens the lunar damping term.
At the same time, Eq.~\eqref{eq:de_dt_moon_energy} still contains an explicit factor of $a$, and $\dot E_{\mathrm{tide}}$ itself depends strongly on $a$ through the tidal forcing. The lunar damping rate is therefore not determined by $e$ alone: it inherits the uncertainty in the Earth--Moon distance history, which in turn depends primarily on tides raised on the Earth.

Taken together, Eqs.~\eqref{eq:de_dt_total}--\eqref{eq:de_dt_moon_energy} suggest that the early eccentricity history was likely governed by competition between Earth-driven excitation and Moon-driven damping, with the latter strongly modulated by the evolving rheology of the LMO itself. In particular, the mechanism proposed here implies that strong tidal dissipation within the LMO should itself damp $e$ and thereby act to reduce the tidal heating that sustains the high-dissipation state. However, the early lunar eccentricity was almost certainly not characterized by a single simple value: its initial post-formation magnitude, solar and resonant perturbations, and the coupled evolution of the Earth--Moon system likely produced a complicated time history. These equations also show that $de/dt$ cannot be interpreted independently of $a(t)$: like the recession problem discussed below, the eccentricity history is fundamentally entangled with the uncertain tidal, spin, and rheological evolution of the early Earth. For this reason, our use of prescribed eccentricity, especially a relatively low fixed baseline value, should be understood as a controlled simplification that isolates the thermal consequences of LMO tidal heating. More realistic orbital histories will require the Earth term, solar forcing, resonant effects, and the Moon's thermal evolution to be treated together in a fully coupled model. We do not solve this coupled problem here, but these equations provide a natural starting point for a future orbit--thermal model in which $a(t)$, $e(t)$, $\Omega_{\mathrm{E}}(t)$, and the thermal state of the LMO are evolved self-consistently.

\subsection*{3.6 Possible feedback of tidal dissipation on Earth--Moon distance evolution}

In the baseline calculations, the semi-major axis evolution is prescribed by Eq.~\eqref{eq:da_dt}, i.e., by tides raised on the Earth with a constant effective ratio $Q_{\mathrm{E}}/k_{2,\mathrm{E}}$. This provides a convenient orbital clock linking Earth--Moon distance to absolute age. However, if tidal dissipation in either body changes substantially during the magma-ocean epoch, that clock may itself be highly non-uniform. A more general recession equation may be written schematically as
\begin{equation}
  \frac{da}{dt}
  \;=\;
  \left(\frac{da}{dt}\right)_{\mathrm{E}}
  \;+\;
  \left(\frac{da}{dt}\right)_{\mathrm{M}}
  \;+\;
  \left(\frac{da}{dt}\right)_{\mathrm{res}},
  \label{eq:da_dt_total}
\end{equation}
where the three terms denote, respectively, tides raised on the Earth, tides raised on the Moon, and additional resonant or secular contributions.

The Earth term should in principle depend on the terrestrial spin state and on the evolving rheology of the early Earth. A convenient generalized form is
\begin{equation}
  \left(\frac{da}{dt}\right)_{\mathrm{E}}
  \;\approx\;
  3 \left(\frac{k_{2,\mathrm{E}}}{Q_{\mathrm{E}}}\right)
    \left(\frac{M_{\mathrm{M}}}{M_{\mathrm{E}}}\right)
    \left(\frac{R_{\mathrm{E}}}{a}\right)^5
    n a\,
    \mathcal{F}_{\mathrm{E}}\!\left(\frac{\Omega_{\mathrm{E}}}{n},e\right),
  \label{eq:da_dt_earth_general}
\end{equation}
where $\mathcal{F}_{\mathrm{E}}$ summarizes the dependence on terrestrial spin and higher-order eccentricity corrections; our adopted baseline of Eq.~\eqref{eq:da_dt} corresponds effectively to $\mathcal{F}_{\mathrm{E}}=1$. If the Earth remained partially molten and weakly dissipative for an extended interval, its effective tidal response could vary strongly with time, potentially causing lunar recession to proceed much more slowly than in the constant-$Q_{\mathrm{E}}/k_{2,\mathrm{E}}$ approximation\textsuperscript{7,10,11}.

The lunar contribution generally acts in the opposite sense for a synchronously rotating Moon, because dissipation inside the Moon removes orbital energy. At leading order in eccentricity, a classical estimate is
\begin{equation}
  \left(\frac{da}{dt}\right)_{\mathrm{M}}
  \;\approx\;
  -\,21
  \left(\frac{k_{2,\mathrm{M}}}{Q_{\mathrm{M}}}\right)
  \left(\frac{M_{\mathrm{E}}}{M_{\mathrm{M}}}\right)
  \left(\frac{R_{\mathrm{M}}}{a}\right)^5
  n a\, e^2,
  \label{eq:da_dt_moon}
\end{equation}
so stronger dissipation inside the LMO would reduce the net recession rate even while increasing the instantaneous heating. In other words, the same high-dissipation state that prolongs the LMO thermally may also slow the outward migration that eventually weakens the tidal forcing. This introduces a potentially important negative feedback on the orbital clock.

This uncertainty matters for both the mechanism proposed here and the late-heating scenario of Nimmo et al.\textsuperscript{17}. In our model, the thermal cliff occurs when orbital expansion has reduced the forcing enough for the system to leave the stable high-dissipation branch; if $a(t)$ grows more slowly than assumed here, the corresponding cliff age would shift later. Such delayed recession, or even temporary stagnation in $a(t)$, would also keep the Moon at smaller Earth--Moon distance for longer and thus sustain stronger tidal forcing. In that sense, slower orbital expansion could partly offset the damping effect of declining eccentricity by helping the LMO remain in a more strongly forced tidal regime, although the net outcome would still depend on the coupled evolution of $a$, $e$, Earth's tides, and lunar rheology. In the LPT framework of Nimmo et al.\textsuperscript{17}, the timing of the heating event likewise depends on when the Moon reaches the characteristic Earth--Moon distance of the dynamical transition. Thus, although the physical interpretations of the two models differ, both rely on an Earth--Moon distance history whose mapping to absolute time remains uncertain.

For this reason, Eq.~\eqref{eq:da_dt} should be viewed as a practical baseline rather than a unique orbital history. A fully self-consistent treatment would evolve $a(t)$ together with $e(t)$, $\Omega_{\mathrm{E}}(t)$, Earth’s tidal response, and the LMO rheology, thereby allowing the thermal and orbital histories to regulate one another directly. We leave that coupled problem for future work.

\subsection*{3.7 Competition between crystal settling and convective re-entrainment}

Another uncertainty relevant to LMO differentiation is whether crystals that form in the interior can settle efficiently through the convecting magma, or whether vigorous convection continually re-entrains them and delays physical separation. This issue is especially important in the present framework because strong LMO tidal dissipation requires a partially molten crystal--melt mixture to remain mechanically coupled over a substantial volume. If crystals and melt were to separate too efficiently, the tidally active partially molten volume would shrink, viscoelastic dissipation would be suppressed, and the high-dissipation state itself could not be maintained. In that sense, crystal segregation is important not only for geochemical differentiation but also for whether tidal heating can be sustained at all\textsuperscript{25,26}.

For a crystal of radius $r$ and density contrast $\Delta \rho \equiv \rho_{\mathrm{cr}}-\rho_{\mathrm{melt}}$, the downward buoyancy-corrected gravitational force is
\begin{equation}
  F_{\mathrm{set}}
  \;=\;
  \frac{4}{3}\pi r^3 \Delta \rho\, g.
  \label{eq:F_set}
\end{equation}
If the relative motion between the crystal and melt is in the Stokes regime, the opposing viscous drag is
\begin{equation}
  F_{\mathrm{drag}}
  \;=\;
  6\pi \eta_{\mathrm{m}} r\, U_{\mathrm{rel}},
  \label{eq:F_drag}
\end{equation}
where $\eta_{\mathrm{m}}$ is the effective melt viscosity and $U_{\mathrm{rel}}$ is the relative crystal--melt velocity. Equating Eqs.~\eqref{eq:F_set} and \eqref{eq:F_drag} gives the classical settling speed
\begin{equation}
  w_{\mathrm{s}}
  \;\approx\;
  \frac{2\,\Delta \rho\, g\, r^2}{9\,\eta_{\mathrm{m}}}.
  \label{eq:w_set}
\end{equation}
Equation~\eqref{eq:w_set} makes clear that settling becomes easier for larger, denser crystals and harder for more viscous melts.

Whether crystals actually segregate, however, depends on the competition between $w_{\mathrm{s}}$ and the characteristic convective velocity of the magma ocean. A convenient order-of-magnitude estimate for that velocity is
\begin{equation}
  U_{\mathrm{conv}}
  \;\sim\;
  \frac{\kappa}{D_m}\,Ra^{1/3}
  \;\sim\;
  \left(
    \frac{\alpha_T g F_{\mathrm{conv}} D_m}{\rho c_{\mathrm{LMO}}}
  \right)^{1/3},
  \label{eq:u_conv}
\end{equation}
where $D_m$ is the LMO thickness, $Ra$ is the Rayleigh number, and
$F_{\mathrm{conv}}=\dot Q_{\mathrm{conv}}/(4\pi R_{\mathrm{M}}^2)$ is the convective heat flux. It is then useful to define a suspension or re-entrainment parameter
\begin{equation}
  \Pi_{\mathrm{sus}}
  \;\equiv\;
  \frac{U_{\mathrm{conv}}}{w_{\mathrm{s}}}.
  \label{eq:Pi_sus}
\end{equation}
If $\Pi_{\mathrm{sus}} \ll 1$, crystals settle faster than convective motions can recycle them and efficient crystal separation is expected. If $\Pi_{\mathrm{sus}} \gtrsim 1$, convective stirring is strong enough to keep crystals suspended or repeatedly re-entrain them, reducing the net settling flux even when individual crystals are negatively buoyant\textsuperscript{26}.

The effects of additional heat sources on crystal transport in the lunar magma ocean have not been explored in detail. In the present framework, however, there is a direct first-order link between tidal heating and crystal transport. Injecting additional heat into the LMO tends to increase $T_m$, reduce viscosity, and sustain a larger convective heat flux. Through Eq.~\eqref{eq:u_conv}, all three effects act to increase $U_{\mathrm{conv}}$, while the associated reduction in $\eta_{\mathrm{m}}$ and changes in melt fraction can modify $w_{\mathrm{s}}$ and the density contrast. In practice, stronger internal heating should generally shift the system toward larger $\Pi_{\mathrm{sus}}$, thereby favoring suspension/re-entrainment over efficient sedimentation. This should apply not only to the sustained LMO tidal heating considered here, but also to any other heat input delivered while the Moon still hosts a magma ocean, including an LPT-style heating episode if it occurred within the LMO lifetime. In that sense, additional heating should, to first order, intensify convection and favor maintenance of a partially molten, crystal--melt mixed state rather than efficient melt--crystal separation. This is consistent with recent slushy lunar magma-ocean models, in which a nearly uniform crystal suspension persists until the crystal fraction reaches a critical concentration of $\phi_c \approx 0.5$--0.6\textsuperscript{27}. Tidal heating may therefore prolong LMO evolution not only by buffering the thermal budget, but also by dynamically hindering the gravitational segregation of newly formed crystals.

We do not attempt to quantify this effect here, because a realistic calculation would require assumptions about crystal size distributions, phase-dependent density contrasts (e.g., mafic cumulates versus buoyant plagioclase), crystal growth rates, and the non-Newtonian rheology of crystal-bearing mush, as well as the magnitude and timing of any additional heat source. Nevertheless, Eqs.~\eqref{eq:F_set}--\eqref{eq:Pi_sus} provide a useful starting point for estimating when the LMO should behave as a well-mixed crystal suspension and when it should transition to efficient cumulate formation.

\clearpage
\section*{Supplementary Figures}

\begin{center}
\includegraphics[width=0.95\textwidth,height=0.82\textheight,keepaspectratio]{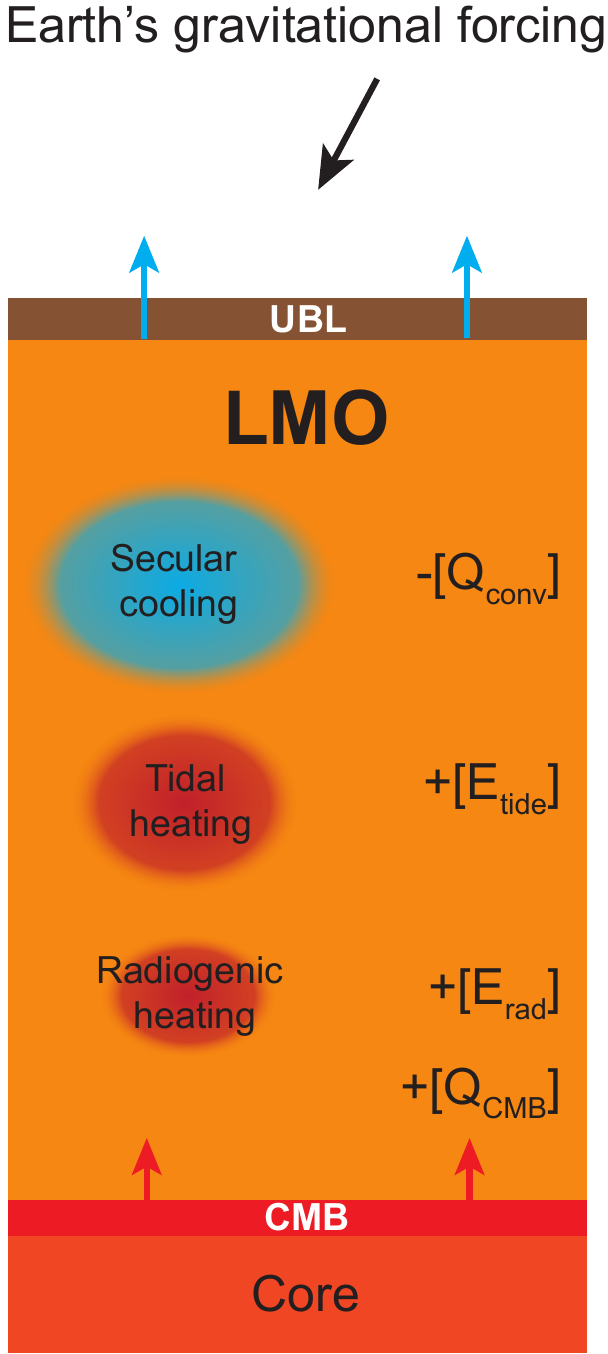}
\end{center}
\noindent \textbf{Supplementary Fig.~S1. Schematic diagram illustrating the thermal energy budget of the LMO.}

\clearpage
\begin{center}
\includegraphics[width=0.95\textwidth,height=0.82\textheight,keepaspectratio]{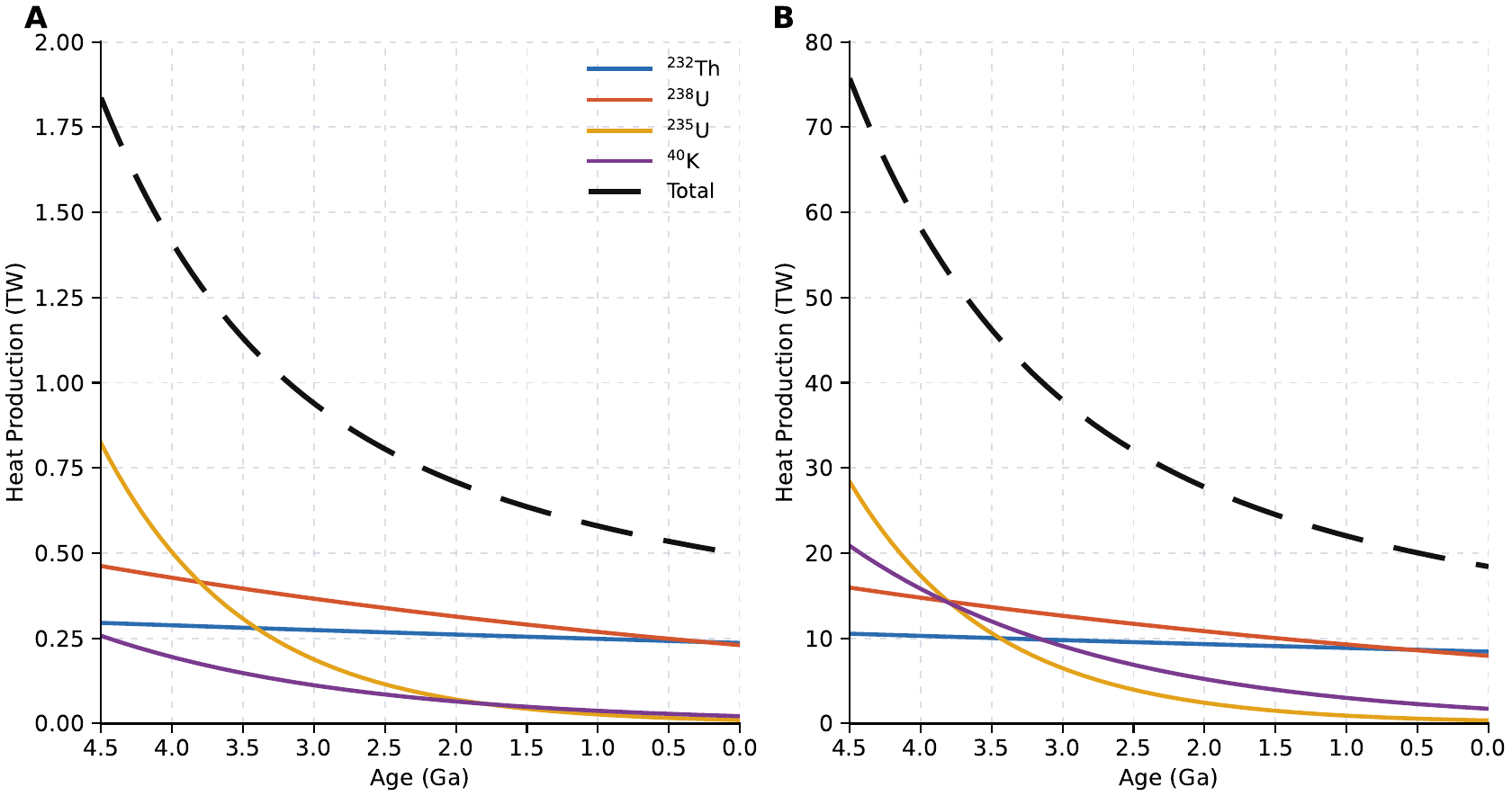}
\end{center}
\noindent \textbf{Supplementary Fig.~S2. Radiogenic heat production rate of the Moon and the Earth.}
Time evolution of radiogenic heat production in the mantles of the Moon and the Earth, based on decay of four key heat-producing isotopes: \textsuperscript{232}Th, \textsuperscript{238}U, \textsuperscript{235}U, and \textsuperscript{40}K. Colored solid lines represent the contribution of each isotope, and the thick black dashed line denotes the total radiogenic heat production. (\textbf{A}) Radiogenic heat in the lunar mantle decreases from an initial value of approximately 1.9~TW at 4.5~Ga to less than 0.5~TW at present, with \textsuperscript{238}U and \textsuperscript{232}Th dominating at later times. (\textbf{B}) Radiogenic heat in the Earth’s mantle starts at $\sim 75$~TW and declines to $\sim 20$~TW today, with \textsuperscript{40}K as the dominant early contributor. The stronger and more sustained radiogenic power in Earth compared to the Moon highlights the latter's limited internal heat budget. The elemental concentrations are from Taylor\textsuperscript{51} and Arevalo Jr.\ et al.\textsuperscript{52}.

\clearpage
\begin{center}
\includegraphics[width=0.95\textwidth,height=0.82\textheight,keepaspectratio]{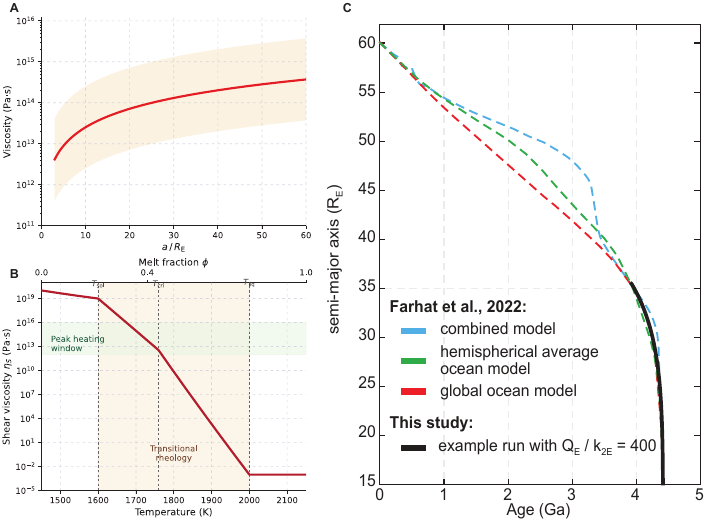}
\end{center}
\noindent \textbf{Supplementary Fig.~S3. Peak tidal viscosity and the adopted LMO viscosity law.}
(\textbf{A}) The red curve shows the peak viscosity $\eta_{\rm peak}(a)$ from the semi-major-axis-dependent peak-heating relation used in the Supplementary Methods. The shaded region illustrates an uncertainty band equal to a factor of ten around $\eta_{\rm peak}$---i.e., $[\eta_{\rm peak}/10,\,10\,\eta_{\rm peak}]$---to account for uncertainties in rheology and forcing frequency. Material parameters are listed in Table~S1. Over $a=3$--$60\,R_{\mathrm{E}}$, $\eta_{\rm peak}$ lies within the transitional-rheology window ($\sim 10^{12}$--$10^{16}$~Pa\,s). (\textbf{B}) Shear viscosity $\eta_S(T)$ from the segmented parameterization adopted in the Supplementary Methods. Vertical dashed lines mark $T_{\rm sol}$, $T_{\rm cri}$, and $T_{\rm liq}$; the upper axis gives the corresponding melt fraction $\phi(T)$. The shaded horizontal band indicates the approximate peak-heating viscosity window, and the shaded vertical band marks the transitional-rheology interval between the solidus and liquidus. (\textbf{C}) Semi-major axis evolution from 4.5 to 4.0 Ga in this work compared to Farhat et al.\textsuperscript{17}.

\clearpage
\begin{center}
\includegraphics[width=0.95\textwidth,height=0.82\textheight,keepaspectratio]{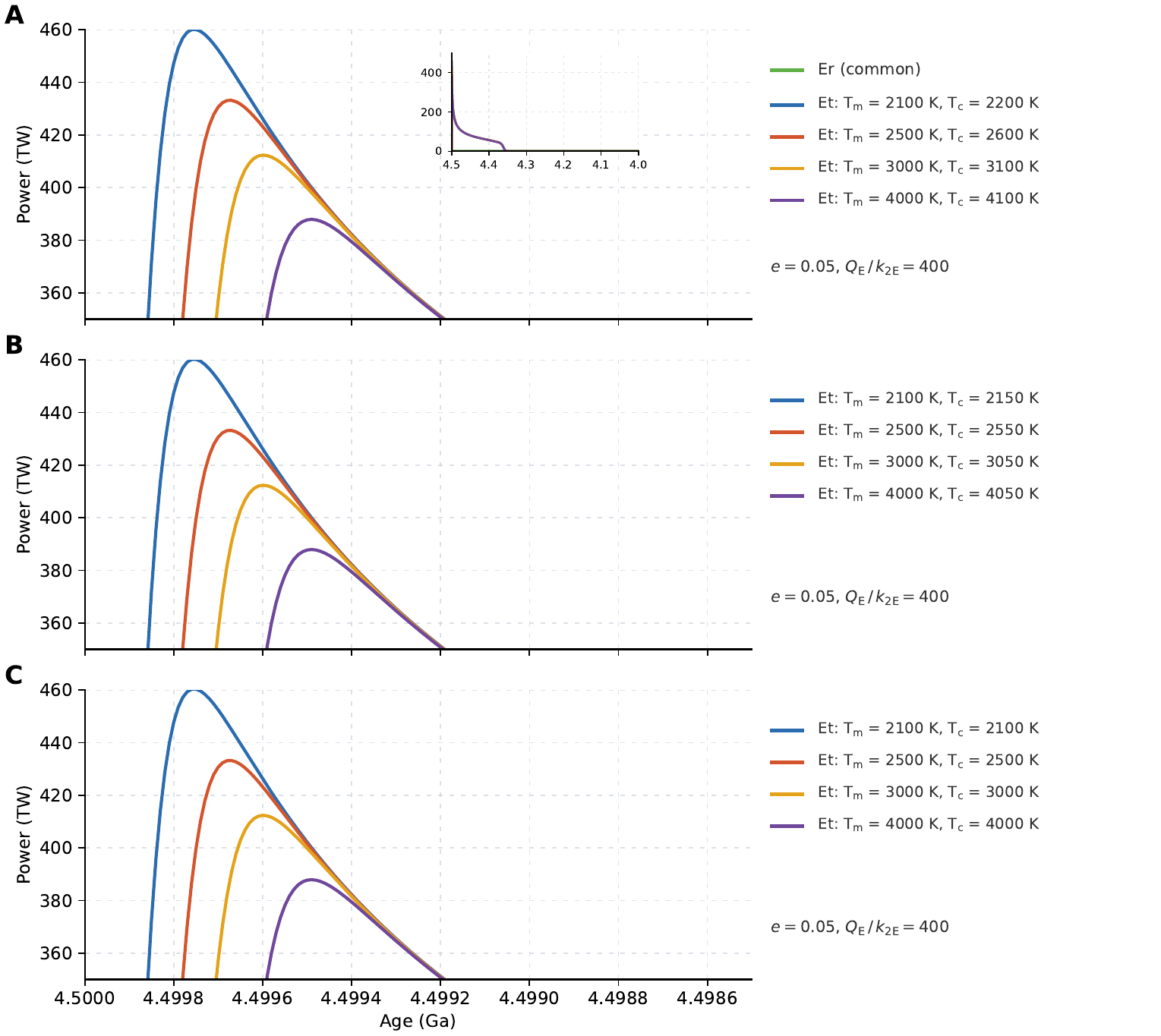}
\end{center}
\noindent \textbf{Supplementary Fig.~S4. Initial LMO and core temperature sensitivity of tidal heating.}
Tidal dissipation power $\dot E_{\mathrm{tide}}$ (colored curves) compared across four initial LMO/core temperature pairs $(T_{m0},T_{c0})$ while holding orbital forcing fixed $(e=0.05,\; Q_{\mathrm{E}}/k_{2\mathrm{E}}=400)$. The common green line is radiogenic heating $\dot E_{\mathrm{rad}}$ (same in all cases). Higher $(T_{m0},T_{c0})$ shift only the short early-time transient---i.e., the entry timing into the transitional-rheology window---and slightly the peak $\dot E_{\mathrm{tide}}$. After the brief start-up, trajectories rapidly converge and the subsequent plateau and the crossing with $\dot E_{\mathrm{rad}}$ (``cliff'' onset) occur at practically the same age and power across all initial-temperature choices. The inset shows the full early spike and the rapid relaxation onto a common track. Panels A--C use different initial core--mantle offsets, but the conclusion is unchanged.

\clearpage
\begin{center}
\includegraphics[width=0.95\textwidth,height=0.82\textheight,keepaspectratio]{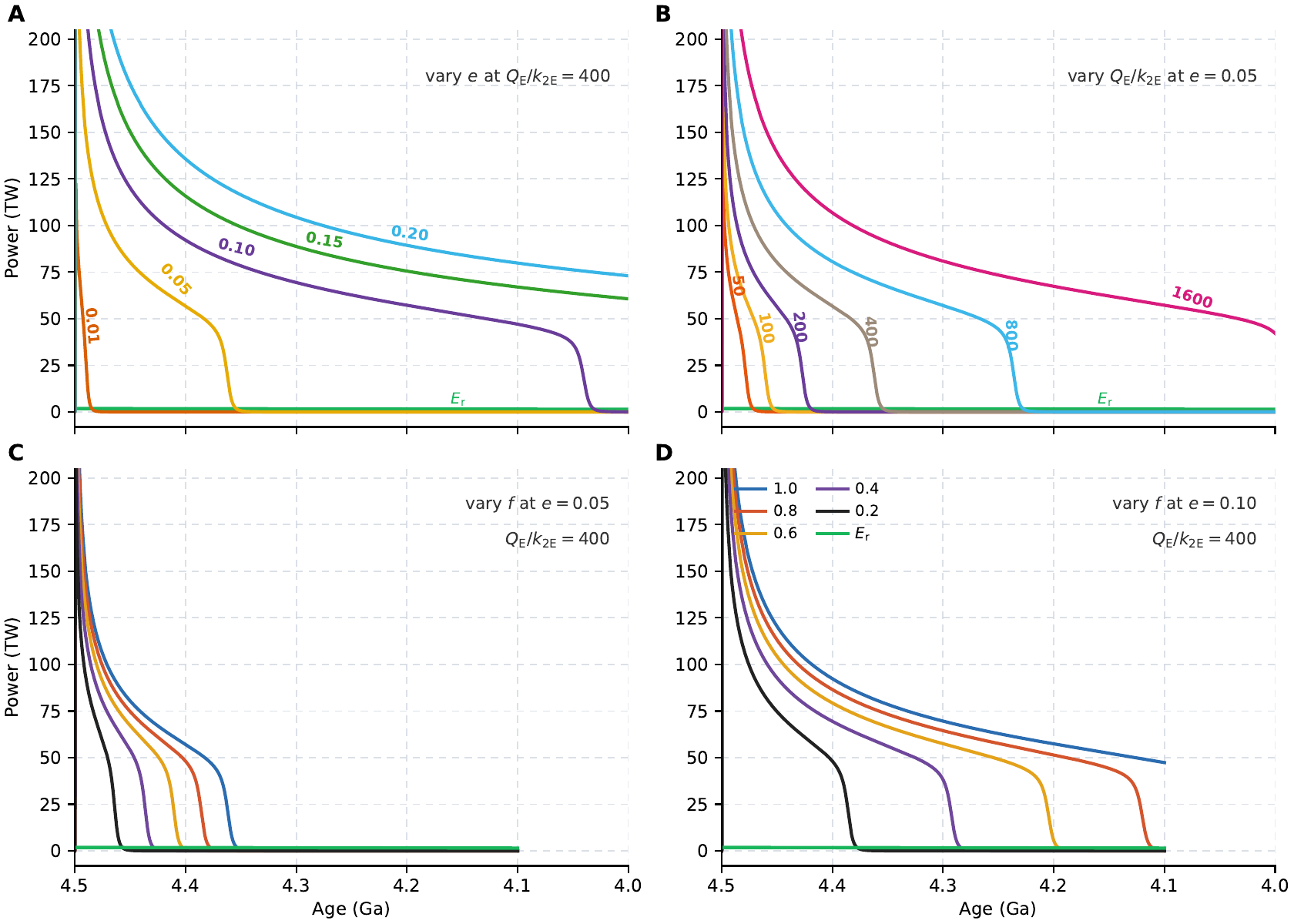}
\end{center}
\noindent \textbf{Supplementary Fig.~S5. Sensitivity of tidal-heating trajectories to orbital forcing and effective participation.}
In all panels, the green curve is the common radiogenic heating $\dot E_{\mathrm{rad}}$. (\textbf{A}) Tidal-heating power $\dot E_{\mathrm{tide}}$ for several eccentricities $e$ at fixed $Q_{\mathrm{E}}/k_{2\mathrm{E}}=400$. (\textbf{B}) $\dot E_{\mathrm{tide}}$ for several $Q_{\mathrm{E}}/k_{2\mathrm{E}}$ values at fixed $e=0.05$; changing $Q_{\mathrm{E}}/k_{2\mathrm{E}}$ modifies the self-consistent $a(t)$ history through the adopted Earth-controlled recession law. (\textbf{C}) $\dot E_{\mathrm{tide}}$ for effective participating fractions $f\in\{1.0,\,0.8,\,0.6,\,0.4,\,0.2\}$ at $e=0.05$ and $Q_{\mathrm{E}}/k_{2\mathrm{E}}=400$. (\textbf{D}) Same as (\textbf{C}) but for $e=0.10$. Larger $e$ or larger $Q_{\mathrm{E}}/k_{2\mathrm{E}}$ sustain higher tidal power for longer and delay the cliff, whereas decreasing $f$ lowers the tidal power and advances the cliff age. At higher eccentricity, the trajectories become less sensitive to $f$ and the cliff shifts to younger ages overall.

\clearpage
\begin{center}
\includegraphics[width=0.95\textwidth,height=0.82\textheight,keepaspectratio]{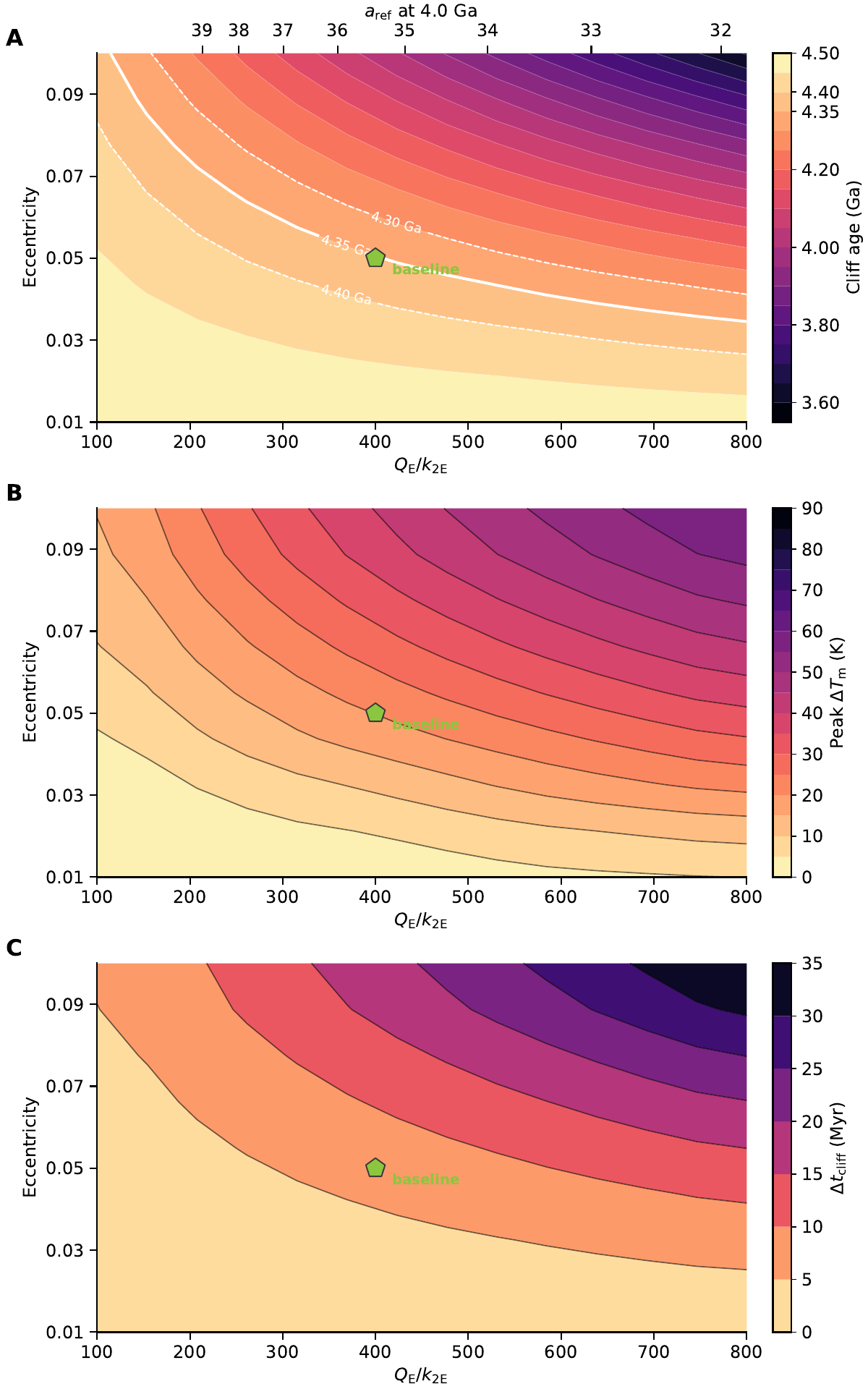}
\end{center}
\noindent \textbf{Supplementary Fig.~S6. Global cliff-age context and near--far asymmetries across the $(e,\;Q_{\mathrm{E}}/k_{2\mathrm{E}})$ space.}
(\textbf{A}) Cliff age predicted by the symmetric baseline model, showing where the terminal thermal collapse occurs in parameter space. (\textbf{B}) Peak near--far mantle temperature contrast, $\max|\Delta T_{\mathrm{m}}|$ (K). (\textbf{C}) Cliff-age difference, $\Delta t_{\mathrm{cliff}}\!\equiv\! t_{\mathrm{cliff}}^{\mathrm{near}}-t_{\mathrm{cliff}}^{\mathrm{far}}$ (Myr); positive values mean the farside reaches the cliff earlier than the nearside. Symbols mark the baseline example case discussed in the main text and supplement.

\clearpage


{\small
\setlength{\LTpre}{0pt}
\setlength{\LTpost}{0pt}
\begin{longtable}{>{\raggedright\arraybackslash}p{0.20\textwidth} >{\raggedright\arraybackslash}p{0.39\textwidth} >{\raggedright\arraybackslash}p{0.29\textwidth}}
\caption{\textbf{List of the parameters used in the model.}}
\label{tab:full_model_parameters_compact}\\
\hline
Symbol & Definition & Value \\
\hline
\endfirsthead
\multicolumn{3}{l}{\textbf{Table \thetable} continued.}\\
\hline
Symbol & Definition & Value \\
\hline
\endhead
\hline
\multicolumn{3}{r}{Continued on next page}\\
\endfoot
\hline
\endlastfoot
$G$ & Gravitational constant & $6.67430\times10^{-11}\ \mathrm{m^{3}\,kg^{-1}\,s^{-2}}$ \\
$\sigma_B$ & Stefan--Boltzmann constant & $5.670374419\times10^{-8}\ \mathrm{W\,m^{-2}\,K^{-4}}$ \\
$R$ & Universal gas constant & $8.314462618\ \mathrm{J\,mol^{-1}\,K^{-1}}$ \\
$M_{\mathrm{E}}$ & Earth mass & $5.972\times10^{24}\ \mathrm{kg}$ \\
$R_{\mathrm{E}}$ & Earth mean radius & $6.371\times10^{6}\ \mathrm{m}$ \\
$M_{\mathrm{M}}$ & Moon mass & $7.342\times10^{22}\ \mathrm{kg}$ \\
$R_{\mathrm{M}}$ & Lunar mean radius & $1.7371\times10^{6}\ \mathrm{m}$ \\
$g$ & Lunar surface gravity & $1.62\ \mathrm{m\,s^{-2}}$ \\
$R_c$ & Lunar core radius & $3.5\times10^{5}\ \mathrm{m}$ \\
$D_m$ & LMO thickness & $1.3871\times10^{6}\ \mathrm{m}$ \\
$\rho$ & LMO bulk density & $3346\ \mathrm{kg\,m^{-3}}$ \\
$c_{\mathrm{LMO}}$ & LMO specific heat capacity & $1200\ \mathrm{J\,kg^{-1}\,K^{-1}}$ \\
$L_{\mathrm{LMO}}$ & Latent heat of fusion & $3.2\times10^{5}\ \mathrm{J\,kg^{-1}}$ \\
$T_s$ & Solidus temperature & $1600\ \mathrm{K}$ \\
$T_l$ & Liquidus temperature & $2000\ \mathrm{K}$ \\
$T_{\mathrm{cri}}$ & Critical rheological temperature & $1760\ \mathrm{K}$ \\
$a_c$ & Nu--Ra prefactor & $0.1$ \\
$Ra_c$ & Critical Rayleigh number (norm.) & $1100$ \\
$\alpha_T$ & Thermal expansivity & $3\times10^{-5}\ \mathrm{K^{-1}}$ \\
$k_m$ & LMO thermal conductivity & $3.75\ \mathrm{W\,m^{-1}\,K^{-1}}$ \\
$\kappa$ & Thermal diffusivity & $1.0\times10^{-6}\ \mathrm{m^{2}\,s^{-1}}$ \\
$\gamma$ & Viscosity-contrast parameter (LBL) & $0.011$ \\
$A$ & Bond albedo & $0.12$ \\
$L_\star$ & Stellar luminosity (Sun) & $3.828\times10^{26}\ \mathrm{W}$ \\
$a_\star$ & Heliocentric distance (1 AU) & $1.496\times10^{11}\ \mathrm{m}$ \\
$\epsilon_v$ & Effective emissivity & $0.9$ \\
$T_{\mathrm{surf},0}$ & Initial surface temperature & $250\ \mathrm{K}$ \\
$T_{m0}$ & Initial LMO temperature & $3000\ \mathrm{K}$ \\
$T_{c0}$ & Initial core temperature & $3100\ \mathrm{K}$ \\
$a_{\mathrm{E}\text{--}\mathrm{M}}$ & Initial Earth--Moon semi-major axis & $3.0\,R_{\mathrm{E}}$ \\
$e$ & Orbital eccentricity & $0.05$ \\
$\mu_0$ & Zero-melt reference modulus & $6.5\times10^{10}\ \mathrm{Pa}$ \\
$\phi_{\mathrm{crit}}$ & Critical melt fraction & $0.40$ \\
$E_{a}$ & Activation energy  & $3.0\times10^{5}\ \mathrm{J\,mol^{-1}}$ \\
$t_{\mathrm{f}}$ & Time since lunar formation & $4.5\ \mathrm{Gyr}$ \\
$H_{232\mathrm{Th}}$ & Specific power (Th-232) & $26.3\times10^{-6}\ \mathrm{W\,kg^{-1}}$ \\
$C_{232\mathrm{Th}}$ & Bulk conc. (Th-232) & $1.25\times10^{-7}\ \mathrm{kg\,kg^{-1}}$ \\
$t_{1/2,232\mathrm{Th}}$ & Half-life (Th-232) & $14.0\ \mathrm{Gyr}$ \\
$\lambda_{232\mathrm{Th}}$ & Decay constant (Th-232) & $1.570\times10^{-18}\ \mathrm{s^{-1}}$ \\
$H_{238\mathrm{U}}$ & Specific power (U-238) & $97.7\times10^{-6}\ \mathrm{W\,kg^{-1}}$ \\
$C_{238\mathrm{U}}$ & Bulk conc. (U-238) & $3.2759\times10^{-8}\ \mathrm{kg\,kg^{-1}}$ \\
$t_{1/2,238\mathrm{U}}$ & Half-life (U-238) & $4.47\ \mathrm{Gyr}$ \\
$\lambda_{238\mathrm{U}}$ & Decay constant (U-238) & $4.917\times10^{-18}\ \mathrm{s^{-1}}$ \\
$H_{235\mathrm{U}}$ & Specific power (U-235) & $574\times10^{-6}\ \mathrm{W\,kg^{-1}}$ \\
$C_{235\mathrm{U}}$ & Bulk conc. (U-235) & $2.376\times10^{-10}\ \mathrm{kg\,kg^{-1}}$ \\
$t_{1/2,235\mathrm{U}}$ & Half-life (U-235) & $0.704\ \mathrm{Gyr}$ \\
$\lambda_{235\mathrm{U}}$ & Decay constant (U-235) & $3.122\times10^{-17}\ \mathrm{s^{-1}}$ \\
$H_{40\mathrm{K}}$ & Specific power (K-40) & $30.4\times10^{-6}\ \mathrm{W\,kg^{-1}}$ \\
$C_{40\mathrm{K}}$ & Bulk conc. (K-40) & $9.711\times10^{-9}\ \mathrm{kg\,kg^{-1}}$ \\
$t_{1/2,40\mathrm{K}}$ & Half-life (K-40) & $1.25\ \mathrm{Gyr}$ \\
$\lambda_{40\mathrm{K}}$ & Decay constant (K-40) & $1.758\times10^{-17}\ \mathrm{s^{-1}}$ \\
$k_{2,\mathrm{E}}$ & Earth degree-2 Love number & $0.3$ \\
$Q_{\mathrm{E}}$ & Earth tidal quality factor & $120$ \\
$M_{c}$ & Core mass & $1.6\times10^{21}\ \mathrm{kg}$ \\
$c_{c}$ & Core specific heat capacity & $840\ \mathrm{J\,kg^{-1}\,K^{-1}}$ \\
$M_{\mathrm{LMO}}$ & LMO mass ($M_{\mathrm{M}}-M_{c}$) & $7.182\times10^{22}\ \mathrm{kg}$ \\
$\mathrm{Nu}_{\max}$ & Nusselt-number cap (UBL) & $10^{5}$ \\
$\eta_{S,\mathrm{sol}}^{\mathrm{ref}}$ & Reference shear viscosity at $T_{\mathrm{sol}}$ & $10^{19}\ \mathrm{Pa\,s}$ \\
$\eta_{S,\mathrm{cri}}^{\mathrm{ref}}$ & Reference shear viscosity at $T_{\mathrm{cri}}$ & $3.5\times10^{12}\ \mathrm{Pa\,s}$ \\
$\eta_{S,\mathrm{liq}}^{\mathrm{ref}}$ & Reference shear viscosity at $T_{\mathrm{liq}}$ & $10^{-3}\ \mathrm{Pa\,s}$ \\
$A_{\phi}$ & Melt-weakening coefficient in viscosity law & 40 \\
$\eta_{\min}$ & Minimum allowed viscosity & $10^{-6}\ \mathrm{Pa\,s}$ \\
$\eta_{\max}$ & Maximum allowed viscosity & $10^{27}\ \mathrm{Pa\,s}$ \\
\end{longtable}
}

\clearpage
\subsection*{Supplementary References}
\begin{enumerate}
\def\labelenumi{\arabic{enumi}.}
\tightlist
\item
  Tyler, R. H., Henning, W. G. \& Hamilton, C. W. Tidal heating in a magma ocean within Jupiter's moon Io. The Astrophysical Journal Supplement Series 218, 22 (2015).
\item
  Renaud, J. P. \& Henning, W. G. Increased tidal dissipation using advanced rheological models: Implications for Io and tidally active exoplanets. The Astrophysical Journal 857, 98 (2018).
\item
  Henning, W. G., O'Connell, R. J. \& Sasselov, D. D. Tidally heated terrestrial exoplanets: viscoelastic response models. The Astrophysical Journal 707, 1000 (2009).
\item
  Nicholls, H. et al.\ Self-limited tidal heating and prolonged magma oceans in the L 98-59 system. Monthly Notices of the Royal Astronomical Society 541, 2566--2584 (2025).
\item
  Efroimsky, M. Tidal dissipation compared to seismic dissipation: In small bodies, Earths, and super-Earths. The Astrophysical Journal 746, 150 (2012).
\item
  Farhat, M., Auclair-Desrotour, P., Bou\'{e}, G. \& Laskar, J. The resonant tidal evolution of the Earth--Moon distance. Astronomy \& Astrophysics 665, L1 (2022).
\item
  Zahnle, K. J., Lupu, R., Dobrovolskis, A. \& Sleep, N. H. The tethered moon. Earth and Planetary Science Letters 427, 74--82 (2015).
\item
  Korenaga, J. Rapid solidification of Earth's magma ocean limits early lunar recession. Icarus 400, 115564 (2023).
\item
  Korenaga, J. Tidal dissipation within Earth's solidifying magma ocean: III. Effects of matrix compaction. Icarus, 116759 (2025c).
\item
  Korenaga, J. Tidal dissipation within Earth's solidifying magma ocean: I. Effects of inertia and lunar orbital eccentricity. Icarus, 116756 (2025a).
\item
  Korenaga, J. Tidal dissipation within Earth's solidifying magma ocean: II. Atmospheric blanketing and its constraint on tidal heating. Icarus, 116743 (2025b).
\item
  Nimmo, F. \& Stevenson, D. J. Influence of early plate tectonics on the thermal evolution and magnetic field of Mars. Journal of Geophysical Research: Planets 105, 11969--11979 (2000).
\item
  Elkins-Tanton, L. T., Burgess, S. \& Yin, Q.-Z. The lunar magma ocean: Reconciling the solidification process with lunar petrology and geochronology. Earth and Planetary Science Letters 304, 326--336 (2011).
\item
  Henning, W. G. et al.\ Increased Lunar Tidal Heating Due to Consideration of Higher Order Terms in Eccentricity and Advanced Rheological Modeling. 55th Lunar and Planetary Science Conference 3040, 1316 (2024).
\item
  Shoji, D. \& Kurita, K. Thermal--orbital coupled tidal heating and habitability of Martian-sized extrasolar planets around M stars. The Astrophysical Journal 789, 3 (2014).
\item
  Segatz, M., Spohn, T., Ross, M. N. \& Schubert, G. Tidal dissipation, surface heat flow, and figure of viscoelastic models of Io. Icarus 75, 187--206 (1988).
\item
  Nimmo, F., Kleine, T. \& Morbidelli, A. Tidally driven remelting around 4.35 billion years ago indicates the Moon is old. Nature 636, 598--602 (2024).
\item
  Murray, C. D. \& Dermott, S. F. Solar System Dynamics (Cambridge University Press, 1999).
\item
  Quillen, A. C., Martini, L. \& Nakajima, M. Near/far side asymmetry in the tidally heated Moon. Icarus 329, 182--196 (2019).
\item
  Park, R. S. et al.\ Thermal asymmetry in the Moon's mantle inferred from monthly tidal response. Nature 641, 1188--1192 (2025).
\item
  Chen, E. M. A. \& Nimmo, F. Tidal dissipation in the lunar magma ocean and its effect on the early evolution of the Earth--Moon system. Icarus 275, 132--142 (2016).
\item
  Peale, S. J. Generalized Cassini's laws. Astronomical Journal 74, 483 (1969).
\item
  Ward, W. R. Tidal friction and generalized Cassini's laws in the solar system. Astronomical Journal 80, 64--70 (1975).
\item
  Garrick-Bethell, I., Wisdom, J. \& Zuber, M. T. Evidence for a past high-eccentricity lunar orbit. Science 313, 652--655 (2006).
\item
  Solomatov, V. Magma oceans and primordial mantle differentiation. In Treatise on Geophysics, 2nd edn, Vol.\ 9, 81--104 (Elsevier, 2015).
\item
  Martin, D. \& Nokes, R. Crystal settling in a vigorously convecting magma chamber. Nature 332, 534--536 (1988).
\item
  Michaut, C. \& Neufeld, J. A. Formation of the Lunar Primary Crust From a Long-Lived Slushy Magma Ocean. Geophysical Research Letters 49, e2021GL095408 (2022).
\item
  Taylor, S. R. Lunar and terrestrial crusts: a contrast in origin and evolution. Physics of the Earth and Planetary Interiors 29, 233--241 (1982).
\item
  Arevalo Jr., R., McDonough, W. F. \& Luong, M. The K/U ratio of the silicate Earth: Insights into mantle composition, structure and thermal evolution. Earth and Planetary Science Letters 278, 361--369 (2009).
\end{enumerate}

\end{document}